\documentclass[twocolumn]{article}
\usepackage{arxiv}

\usepackage[utf8]{inputenc} % allow utf-8 input
\usepackage[T1]{fontenc}    % use 8-bit T1 fonts
\usepackage{hyperref}       % hyperlinks
\usepackage{url}            % simple URL typesetting
\usepackage{booktabs}       % professional-quality tables
\usepackage{amsfonts}       % blackboard math symbols
\usepackage{nicefrac}       % compact symbols for 1/2, etc.
\usepackage{microtype}      % microtypography
\usepackage{doi}
\usepackage{makecell}
\usepackage{amsmath}
\usepackage[numbers]{natbib}
\usepackage[capitalise,noabbrev]{cleveref}
\usepackage{graphicx}
\usepackage{cuted}

\graphicspath{ {./images/} }
\usepackage{xcolor}
\usepackage[acronym]{glossaries}
\usepackage{fancyhdr} % Used to add line in header
\glsdisablehyper  % Use this to remove all the hyperlinks on every acronym

\newcommand{\affit}[1]{$^{\mathrm{\textnormal{\textit{#1}}}}$}
\newacronym{ecmwf}{ECMWF}{European Centre for Medium-Range Weather Forecasts}
\newacronym{ddms}{DDMs}{Data-driven models}
\newacronym{ddm}{DDM}{data-driven model}
\newacronym{eps}{EPS}{ensemble prediction system}
\newacronym{epss}{EPSs}{ensemble prediction systems}
\newacronym{metno}{MET Norway}{Norwegian Meteorological Institute}
\newacronym{metcoop}{MetCoOp}{Meteorological Co-operation on Operational \acrshort{nwp}}
\newacronym{nwp}{NWP}{numerical weather prediction}
\newacronym{synop}{SYNOP}{surface synoptic observations}
\newacronym{nmhses}{NMHSes}{national meteorological and hydrological services}
\newacronym{gnns}{GNNs}{graph neural networks}
\newacronym{gnn}{GNN}{graph neural network}
\newacronym{gpu}{GPU}{graphical processing unit}
\newacronym{gpus}{GPUs}{graphical processing units}
\newacronym{lam}{LAM}{limited area model}
\newacronym{lams}{LAMs}{limited area models}
\newacronym{tdp}{TDP}{thermal design power}

\newacronym{t}{T}{2-meter temperature}
\newacronym{ws}{WS}{10-meter wind speed}
\newacronym{p6h}{P6h}{6-hour precipitation accumulation}
\newacronym{mslp}{MSLP}{mean sea-level pressure}
\newacronym{tmin}{Tmin}{daily minimum 2-meter temperature}
\newacronym{tmax}{Tmax}{daily maximum 2-meter temperature}
\newacronym{wsmax}{WSmax}{daily maximum wind speed}
\newacronym{p24h}{P24h}{24-hour precipitation accumulation}

\newacronym{ifs}{IFS}{Integrated Forecast System}
\newacronym{meps}{MEPS}{MetCoOp Ensemble Prediction System}
\newacronym{aifs}{AIFS}{Artificial Intelligence Forecasting System}
\newacronym{bris}{Bris CRPS-FFT}{Bris CRPS-FFT}

\newacronym{fft}{FFT}{fast Fourier transform}
\newacronym{relu}{ReLU}{rectified linear unit}
\newacronym{mlp}{MLP}{multilayer perceptron}
\newacronym{nlp}{NLP}{Natural Language Processing}

\newacronym{crps}{CRPS}{Continuous Ranked Probability Score}
\newacronym{fcrps}{fCRPS}{Fair Continuous Ranked Probability Score}
\newacronym{ssr}{SSR}{spread-skill ratio}
\newacronym{mse}{MSE}{mean squared error}
\newacronym{mae}{MAE}{mean absolute error}
\newacronym{rmse}{RMSE}{root mean square error}
\newacronym{ets}{ETS}{equitable threat score}
\newacronym{fss}{FSS}{fractions skill score}
\newacronym{dct}{DCT}{discrete cosine transform}
\newacronym{bss}{BSS}{Brier skill score}
\newacronym{qq}{QQ}{quantile-quantile}

\title{High-Resolution Probabilistic Data-Driven Weather Modeling with a Stretched Grid}

\author{
 Even Marius Nordhagen \affit{a*}\textsuperscript{\textdagger} \\
  \And
 Håvard Homleid Haugen \affit{a}\textsuperscript{\textdagger} \\ 
  \And
 Magnus Sikora Ingstad \affit{a} \\
  \And
 Aram Farhad Shafiq Salihi \affit{a} \\
  \And
 Thomas Nils Nipen \affit{a} \\
  \And
 Ivar Ambjørn Seierstad \affit{a} \\
  \And
 Inger-Lise Frogner \affit{a} \\
  \And
 Mariana C. A. Clare \affit{c} \\
  \And
 Simon Lang \affit{b} \\
  \And
 Matthew Chantry \affit{b} \\
  \And
 Peter Dueben \affit{c} \\
  \And
 Jørn Kristiansen \affit{a} \\
}

\begin{document}

\twocolumn[
\begin{@twocolumnfalse}
\maketitle
\bigskip
\begin{abstract}

We present a probabilistic data-driven weather model providing ensembles of high spatial resolution realizations of 87 variables at arbitrary ensemble size and forecast length.
The model uses a global stretched grid, dedicating 2.5 km resolution to our Nordic region of interest and 31 km resolution elsewhere, with 6-hour temporal resolution.
Unique ensemble members are generated by a stochastic model architecture, and we train it using a loss function based on the Continuous Ranked Probability Score (CRPS) evaluated in grid-point and spectral space.
The spectral loss component is shown to be necessary to create fields that are spatially coherent, which is not the case when training with mean-squared error loss, nor CRPS in grid-point space only.
We evaluate the forecasts against observations from surface weather stations and compare them to high-resolution operational numerical weather prediction forecasts from the MetCoOp Ensemble Prediction System (MEPS).
The model shows lower CRPS than MEPS for 2 m temperature and mean sea-level pressure, with average improvements of 13\% and 10\%, respectively, while differences for wind speed and precipitation are smaller.
For Storm Dave, the model captures the location and structure of strong-wind systems, but underestimates the peak winds.

\end{abstract}

\bigskip
\vspace{1.5cm}
\end{@twocolumnfalse}
]

\footnotetext[1]{Norwegian Meteorological Institute, Oslo, Norway}
\footnotetext[2]{ECMWF, Reading, UK}
\footnotetext[3]{ECMWF, Bonn, Germany}
% European Centre for Medium-Range Weather Forecasts
\footnotetext{\textsuperscript{\textdagger}Equal contribution}
\footnotetext{*Corresponding author: \texttt{even.nordhagen@gmail.com}}

%%%%%%%%%%%%%%
% INTRODUCTION %
%%%%%%%%%%%%%%

\begin{figure*}[t]
    \centering
    \includegraphics[width=\textwidth]{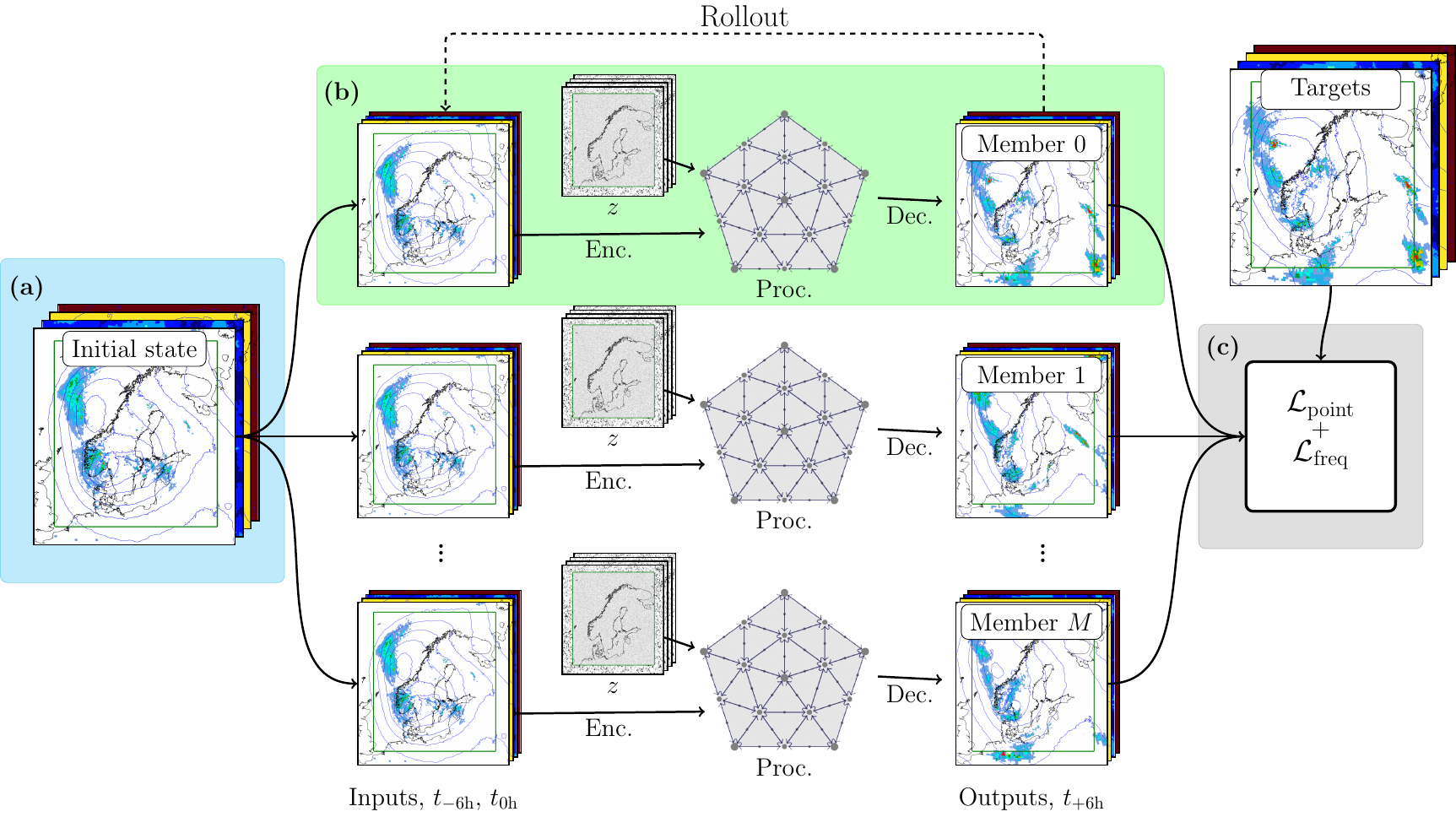}
    \caption{Schematic diagram of the ensemble training. (a) The preceding ($t_{-6\mathrm{h}}$) and the current ($t_{0\mathrm{h}}$) states from MEPS and ERA5 are used as inputs to the model. The inputs are replicated $M$ times, where $M$ is the number of ensemble members during training. (b) The model has an encoder-processor-decoder architecture, where noise ($z$) is injected into the latent space for stochasticity. (c) The predictions are evaluated against targets using a point-wise and spectral training objective.}
    \label{fig:model}
\end{figure*}  

\section{Introduction}\label{sec:intro}

\acrfull{ddms} have recently been shown to achieve comparable performance to conventional \acrfull{nwp} systems across a range of forecast variables when evaluated using \acrfull{mse}, while requiring only a fraction of the computational cost. Many of these models, including FourCastNet \citep{pathak2022}, PanguWeather \citep{bi2023}, FuXi \citep{chen2024}, and the \acrfull{aifs} \citep{lang2024}, are deterministic neural network-based  models trained with a \acrshort{mse} loss function. More recently, probabilistic approaches based on proper scoring rules or generative modeling have been introduced, including AIFS-CRPS \citep{lang2026}, FGN \citep{alet2025}, FourCastNet3 \citep{bonev2025}, and GenCast \citep{price2025}. These developments reflect a growing recognition of the limitations of deterministic training objectives such as \acrshort{mse}. In particular, \acrshort{mse} is affected by the double penalty problem \citep{ebert2000,zingerle2008,subich2025}: when a model predicts an event slightly displaced from its observed location, it is penalized both for the false prediction and for the missed true event. Minimizing \acrshort{mse} therefore encourage models to smooth less predictable features, particularly extreme events.

Addressing this challenge requires a training objective that is not based purely on the \acrshort{mse}. One approach is to retain the deterministic framing, but modify the loss function to encourage sharper spatial fields \citep{subich2025,xu2024}. Another alternative is to adopt probabilistic models that explicitly represent forecast uncertainty, which is also essential to predict extreme weather events. In data-driven weather prediction, these have predominantly relied on proper score-based and diffusion-based methods.

Proper score-based approaches \citep{kochkov2024,lang2026,alet2025,bonev2025,schillinger2025} directly optimize proper scoring rules such as \acrfull{crps} or the energy score, encouraging a reliable spread–skill relationship \citep{gneiting2007}. Stochasticity is incorporated into the model to generate distinct ensemble members. Because the architecture remains close to its deterministic counterpart, mainly the loss function is changed and noise is injected into the model, score-based models have nearly identical computational cost per member as deterministic models.

Diffusion-based models \citep{price2025,alexe2024a,larsson2025} generally use similar architectures to deterministic networks, but are conditioned on varying noise levels. They are then trained to iteratively remove noise from input fields until sharp forecasts emerge. They are known for producing high-fidelity outputs, and because noise serves as their input, they are inherently stochastic and naturally suited for generating ensembles. However, each forecast typically requires 10–100 denoising steps, making diffusion models substantially more computationally expensive than deterministic and proper score-based approaches.

Global data-driven models typically operate at 0.25° resolution. 
Regional models can achieve significantly higher resolution, by focusing the model's complexity on a small area. 
Regional data-driven models have recently been developed to exploit kilometer-scale forecasting, typically using either limited-area models (LAMs) \citep{oskarsson2024,larsson2025,adamov2025}, where a model receives boundary conditions from an external model, or stretched-grid models \citep{nipen2026,bano-medina2025}, which are global models with higher resolution over an area of interest and lower resolution elsewhere. For high-resolution modeling, the smoothing effect of the \acrshort{mse} loss is significant. As will be demonstrated, applying a CRPS approach directly to the high-resolution case creates spatial fields that are spatially incoherent. We therefore further modify the CRPS approach by introducing a spectral term in the loss function that encourages correct spread at different spatial scales. The spectral term computes the CRPS of the Fourier coefficients obtained via the \acrfull{fft}. We refer to this model as \textit{\acrshort{bris}}. To our knowledge, \acrshort{bris} is the first ensemble \acrshort{ddm} capable of realistically simulating weather at kilometer scale. We compare \acrshort{bris} to a model trained with CRPS applied in grid-point space only, and the deterministic \acrshort{mse}-based model in \citep{nipen2026}, here referred to as \textit{Bris CRPS} and \textit{Bris MSE}, respectively.

% outline
This paper is organized as follows: The model architecture and training approach are described in Sec.~\ref{sec:method}. Sec.~\ref{sec:datasets} discusses the datasets used to train the model and perform inference. In Sec.~\ref{sec:results}, we verify the model against observations from weather stations, evaluating the characteristics of individual ensemble members as well as the probabilistic capabilities of the ensemble as a whole. Finally, conclusions are presented in Sec.~\ref{sec:conclusion}.

\begin{table}[t]
\caption{Model configurations, key properties and datasets used in the different training stages.\\}
\centering
\begin{tabular}{@{}l|lll@{}}
\toprule
& Stage A & Stage B & Stage C \\
\midrule
Global dataset & ERA5 & IFS & IFS \\
Global resolution & 31 km & 31 km & 31 km \\
Global refinement & 7 & 7 & 7 \\
Regional dataset & -- & -- & MEPS \\
Regional resolution & -- & -- & 2.5 km \\
Regional refinement & -- & -- & 10 \\
Number of grid nodes & 540 k & 540 k & 1.3 M \\
Number of mesh nodes & 164 k & 164 k & 284 k \\
Total number of edges & 4.9 M & 4.9 M & 10.3 M \\
Training years & 43 & 6 & 5.2 \\
Ensemble size & 4 & 4 & 2 \\
%Training years & 1979-2021 & 2016-2021 & 2020-2025 \\
\bottomrule
\end{tabular}
\label{tab:stages}
\end{table}

\section{Methodology}\label{sec:method}

\subsection{Model architecture}\label{sec:architecture}

The model uses \acrfull{gnns}, which were first introduced to weather prediction by \cite{keisler2022}, and frequently used in other models \citep[e.g.][]{lang2024,lang2026}. \acrshort{gnns} consist of nodes, which are used to represent the atmosphere for a specific spatial location, and edges, which represent how nodes communicate information. 
This flexible architecture supports arbitrary grid configurations. We follow the stretched-grid approach developed in \citep{nipen2026}, using regional analyses from the \acrfull{meps} \citep{frogner2019a,eresmaa2026,muller2017} at 2.5 km resolution over the Nordic region and IFS data interpolated to 31 km resolution elsewhere on the globe.

The two input datasets are passed through an \textit{encoder} responsible for encoding to a hidden mesh at lower spatial resolution than the input grid, but with a greater number of features. 
The hidden mesh has variable spacing that scales with the local resolution of the input grid. 
This latent representation is then passed through a \textit{processor}, consisting of a number of message passing steps that allow the model to communicate information in space. Finally, the latent representation is passed through a \textit{decoder}, responsible for decoding back to the original input grid. This process generates a 6-hour forecast, and the process can be repeated autoregressively to generate arbitrarily long forecasts.

We have used the same architecture settings as in \cite{nipen2026, lang2026}, such as the embedding dimension (1024), number of attention heads (8) and number of message-passing steps (16), resulting in a similar number of trainable parameters (229 million). The graph architecture is similar to that of \cite{nipen2026}, with an icosahedral hidden mesh refined 7 times globally and 10 times over the regional domain, and with 12 encoder connections per mesh node and 3 decoder connections per grid node. This gives $N_{\mathrm{mesh}}=284,000$ hidden mesh nodes. In total, the model has 1.3 million input/output grid points.

The key novelty of our model is that it provides ensemble forecasts at kilometre scale. To represent stochastic processes in the atmosphere and to ensure inherently different ensemble members for an arbitrary ensemble size, the model needs to be stochastic. This is done by injecting noise into the latent space through a \acrfull{mlp} and conditional layer normalization, following AIFS-CRPS \citep{lang2026} (Fig.~\ref{fig:model}). The \acrshort{mlp} takes in Gaussian noise with dimensions (4, $N_{\mathrm{mesh}}$), allowing the model to shape the noise according to the fields. The noise injector also takes information about the forecast step, allowing the model to incorporate different noise at different forecast lengths.

In order to generate skillful ensemble forecasts, the model is trained with a loss based on \acrfull{crps}, the almost-fair \acrshort{crps} \citep{lang2026}. To calculate the loss, for each sample an ensemble is required during training. Since \acrshort{crps} evaluates the cumulative distribution function at individual grid points to compute the loss, the training ensemble size can be as small as 2 \citep{lang2026}. During inference, an arbitrary number of members can be generated.

\begin{table*}[t]
\caption{Training configuration of the different training stages and substages. Learning rate here refers to the effective learning rate used in the parameter update.\\}
\centering
\begin{tabular}{@{}ll|llll|ll|llll@{}}
\toprule
&Stage && A1 & A2 & A3 && B && C1 & C2 & \\
\midrule
&Rollout && 1 & 2 & 3--6 && 6 && 1 & 2--6 \\
&Iterations && 300 k & 60 k & 1,967 && 2,720 && 40 k & 2 k \\
&Learning rate && $1\cdot 10^{-3}$ & $1\cdot 10^{-5}$ & $1\cdot 10^{-5}$ && $1\cdot 10^{-5}$ && $2\cdot 10^{-4}$ & $6\cdot 10^{-5}$ \\
&Warm up iterations && 1,000 & 1,000 & 100 && 100 && 1,000 & 100 \\
&Ensemble size && 4 & 4 & 4 && 4 && 2 & 2 \\
%&Training speed (s/iter) && 4--5 & 6--7 & 12--25 && 23--25 && 7--9 & 16--50\\
%&Training period (years) && 43 & 43 & 43 && 6 && 5.2 & 5.2 \\
%&Batch size && 16 & 16 & 16 && 16 && 16 & 16 \\
%&GPUs per model && 4 & 4 & 4 && 4 && 4 & 4 \\
%&GPUs && 256 & 256 & 256 && 256 && 128 & 128 \\
%&Total GPU-hours && 89,000 & 30,000 & 10,000 && 3,500 && 12,000 & 11,000 \\
\bottomrule
\end{tabular}
\label{tab:substages}
\end{table*}

\subsection{Training objective} \label{sec:loss}

The \acrfull{crps} loss depends only on the marginal distributions of the ensemble at each location and therefore is not scale-aware \citep{lang2025}.
Members can therefore appear to be spatially incoherent when the model generates too much or too little variability at a certain scale.

Different approaches have been proposed to address this issue. One strategy is to decompose the fields into different spatial scales and evaluate \acrshort{crps} separately for each scale \citep{lang2025}. Another is to compute \acrshort{crps} both in the grid-point space and spectral space \citep{kochkov2024,bonev2025}, which is the approach we are using here.

The training objective consists of a global and a regional component, similar to \citep{nipen2026}. The global component, evaluated on the 31 km horizontal resolution grid, is the \acrshort{crps} computed point-wise in space. The regional component, evaluated at 2.5 km horizontal resolution, combines a grid-point space \acrshort{crps} loss with a spectral \acrshort{crps} loss.

Mathematically, the training objective can be expressed as:
\begin{equation*}
    \mathcal{L}(\{x\},y)=\sum_v\sum_tw_v\bigg[\mathcal{L}_{\mathrm{g}}(\{x_{\mathrm{g}}\},y_{\mathrm{g}})+\lambda_{\mathrm{r}}\mathcal{L}_{\mathrm{r}}(\{x_{\mathrm{r}}\},y_{\mathrm{r}})\bigg],
\end{equation*}
where $\mathcal{L}_{\mathrm{g}}$ and $\mathcal{L}_{\mathrm{r}}$ denote the global and regional loss terms, respectively. Here, $\{x\}$ denotes the ensemble predictions, $y$ the targets, $v$ the variables, $t$ the rollout length, and $w_v$ the variable-specific scalings, taken from \citep{lang2024}. The predictions are divided into global and regional domains $\{x\}=\{x_{\mathrm{g}}\}\cup\{x_{\mathrm{r}}\}$, with the same split applied to the targets. Note that the regional domain in our case is on a regular Cartesian grid, which is convenient for applying transforms. The relative importance of the regional loss is determined by $\lambda_{\mathrm{r}}$.

The global and regional loss terms are, respectively:
\begin{equation}
    \begin{aligned}
        \mathcal{L}_{\mathrm{g}}(\{x\},y)&=\mathcal{L}_{\mathrm{point}}(\{x\},y), \\
        \mathcal{L}_{\mathrm{r}}(\{x\},y)&=\mathcal{L}_{\mathrm{point}}(\{x\},y)+\lambda_{\mathrm{f}}\mathcal{L}_{\mathrm{freq}}(\{x\},y),
    \end{aligned}
\end{equation}

here in its kernel representation \cite{leutbecher2019}:

\begin{strip}
\begin{equation}
    \mathcal{L}_{\mathrm{point}}\left(\{x\},y\right)=\sum_pw_p\left[\frac{1}{M}\sum_{i=1}^M|x_i-y|-\frac{1-\varepsilon}{2M(M-1)}\sum_{i=1}^M\sum_{j=1}^M|x_i-x_j|\right].
    \label{eq:afcrps}
\end{equation}
\end{strip}
Here, $p$ denotes a spatial node and $w_p$ the corresponding node weight. $M$ is the ensemble size, and $\varepsilon$ determines the degree of \acrshort{crps} ``fairness'': $\varepsilon=0$ corresponds to a fully fair \acrshort{crps}, while $\varepsilon=1/M$ yields the conventional \acrshort{crps}. Values close to zero produce an almost-fair loss.

For the spectral loss component, $\mathcal{L}_{\mathrm{freq}}$, we compute the two-dimensional \acrfull{fft} of each target field and of each ensemble member. The almost-fair \acrshort{crps} loss is then applied directly to the resulting Fourier coefficients, which are complex-valued and therefore retain both phase and magnitude information. In the \acrshort{crps} computation, absolute values are interpreted as complex moduli, ensuring that the resulting loss is real:
\begin{equation}
    \mathcal{L}_{\mathrm{freq}}\big(\{x\},y\big)=\mathcal{L}_{\mathrm{point}}\big(\{\mathrm{FFT}(x)\},\mathrm{FFT}(y)\big),
\end{equation}
where $\mathrm{FFT}(x)$ denotes the Fourier coefficients of field $x$. The fields are reshaped to rectangular grids before applying the \acrshort{fft}. The spectral term is weighted with a factor $\lambda_{\mathrm{f}}$ relative to the point-wise term. We perform no frequency weighting, i.e., $w_p=1$ for the spectral term. Also, note that the variable weights $w_v$ are the same for the point-wise and spectral terms, but the sharp surface variables (e.g., 6-hour accumulated precipitation, wind speed at 10 m, temperature at 2 m) will have a higher natural magnitude in Fourier space.

Although these additional terms increase the complexity of the loss computation, we did not observe a noticeable increase in overall computation time. %The computational cost remains dominated by other factors, such as model throughput and gradient updates.

\begin{table*}
 \caption{List of variables used as input and output in the model. Pressure fields are available on 50 hPa, 100 hPa, 150 hPa, 200 hPa, 250 hPa, 300 hPa, 400 hPa, 500 hPa, 700 hPa, 800 hPa, 850 hPa, 925 hPa, 1000 hPa levels. Forcing fields are input fields only.}
  \centering
  \begin{tabular}{lll}
    \toprule
    Pressure level fields & Single level fields & Forcing fields \\
    \midrule
    Geopotential height & Skin/sea-surface temperature & Solar insolation \\
    Temperature & 2 m temperature & Sine of Julian day \\
    Specific humidity & 2 m dew point temperature & Sine of latitude \\
    Wind speed u component & 10 m wind speed u component & Sine of local time \\
    Wind speed v component & 10 m wind speed v component & Sine of longitude \\
    Wind speed w component & mean sea-level pressure & Cosine of Julian day \\
    & Surface air pressure & Cosine of longitude \\
    & Total column integrated water & Cosine of latitude \\
    & 6-hour accumulated precipitation*\textdagger & Cosine of local time \\
    & Low cloud coverage\textdaggerdbl & Land sea mask \\ 
    & Medium cloud coverage\textdaggerdbl & Surface geopotential height \\
    & High cloud coverage\textdaggerdbl \\
    & Total cloud coverage\textdaggerdbl \\
    & Surface solar radiation downwards* \\                              
    & Surface thermal radiation downwards* \\                              
    \bottomrule
    \multicolumn{3}{l}{* diagnostic variable} \\
    \multicolumn{3}{l}{\textdagger\, \acrfull{relu} bounding applied} \\
    \multicolumn{3}{l}{\textdaggerdbl\, hard tanh bounding applied with range 0 to 1}
  \end{tabular}
  \label{tab:vars}
\end{table*}

\subsection{Training procedure} \label{sec:procedure}
We follow a stage-wise training procedure with transfer learning between the different stages, separated by datasets and training periods. An overview of the stages can be seen in Table~\ref{tab:stages}.

In stage A, we pre-train a global model on ERA5 with its native N320 grid ($\sim 31\, \mathrm{km}$ horizontal resolution). This stage is important for the model to learn general weather patterns, necessary since the length of the regional dataset is limited. This stage follows the training approach described in \citep{lang2026}, where the model is trained for 300,000 iterations on 6-hour predictions (stage A1), 60,000 iterations with a 12-hour rollout (stage A2), and finally 1,967 iterations (1 epoch) for each rollout length from 18 to 36 hours (stage A3) to learn multi-step properties. We stopped after 6 rollout steps (36 hours), unlike AIFS-CRPS which went to 12 rollout steps, because the effect of longer rollout is minor for short-term forecasting, see Appendix~\ref{sec:rollout}. 

In stage B, the model is fine-tuned on operational analysis, \acrfull{ifs}, interpolated to the N320 grid, continuing with a 36-hour rollout (6 rollout steps) for 2,720 iterations (5 epochs). \acrshort{ifs} has higher native resolution than ERA5, with fields that exhibit greater variability. This stage is responsible for preparing the model for operational use.

Stage C is where we introduce the stretched grid, using the \acrshort{ifs} analysis globally and the \acrshort{meps} analysis over the Nordic domain. In stage C1, we apply learning rate $2\cdot10^{-4}$ for 40,000 iterations and train with regional weighting $\lambda_{\mathrm{r}}=0.23$ and spectral loss weighting $\lambda_{\mathrm{f}}=0.1$. In stage C2, we perform 2,000 iterations at each rollout length from 12 to 36 hours. These hyperparameters were picked from hyperparameter searches, detailed in Appendix~\ref{sec:hyperparameter}, but we increased the number of iterations proportional to the number of samples in the training data. The training configuration of all substages is presented in Table~\ref{tab:substages}.

We have also trained a model without a spectral term (effectively setting $\lambda_{\mathrm{f}}=0$) as a reference for visual properties. For that model, we start from the pre-trained model from stage B, and run stage C1 only.

An ensemble size of $M=4$ is used during pre-training (stages A and B), while $M=2$ is used for stage C. We keep \acrshort{crps} ``fairness'' $\varepsilon=0.05/M$ for all runs. 
The AdamW optimizer \citep{loshchilov2019} is used with the default hyperparameters from that paper. A cosine learning-rate schedule is applied, going from the starting learning rate to zero during training and restarting between each substage. We also apply a warm-up period, where the learning rate increases linearly until it reaches the starting learning rate, with 1,000 warm-up steps for stages A1, A2 and C1, and 100 warm-up steps for stages B and C2.

The effect of some of the key hyperparameters are discussed in detail in Appendix~\ref{sec:hyperparameter}.

\subsection{Computational details}\label{sec:computational}
The experiments were carried out on the Leonardo supercomputer \citep{turisini2024}. Each node consists of 4 NVIDIA A100 \acrfull{gpus} equipped with 64 GB GPU memory. The Anemoi framework \citep{lang2024,nipen2026,wijnands2026}, implemented in PyTorch \citep{paszke2019}, is used to build the model, graphs, datasets, and the training pipeline. It supports several complementary strategies for distributing training across multiple \acrshort{gpus}, as described in \cite{lang2026}. Data parallelism enables mini-batches to be processed in parallel across multiple devices, while context parallelism shards the model input grids, processor mesh, and graph edges across devices. In addition, the ensemble dimension can be sharded by assigning different ensemble members of an ensemble group to separate model groups. Here, a model group comprises the devices that together form one sharded model instance, while an ensemble group comprises the model groups that process the members of the same training sample. In our experiments, these sharding strategies are essential to ensure that memory use does not exceed the capacity of any individual \acrshort{gpu}.

In training, we consequently shard the model across $N=4$ \acrshort{gpus} to minimize the communication overhead, and run batches and members in parallel across nodes. The number of \acrshort{gpus} requested during training is therefore $N_{\mathrm{GPUs}}=M\times B\times N$, with $M$ and $B$ as the ensemble and batch size, respectively. We use batch size $B=16$, like in \citep{lang2026}. In stages A and B, the ensemble size is $M=4$, requiring 256 \acrshort{gpus}. In stage C, we use $M=2$, requiring 128 \acrshort{gpus}\footnote{For the longest rollout steps (30 and 36-hour), we had to shard the model across $N=8$ \acrshort{gpus} due to high memory pressure, requiring 256 \acrshort{gpus} in total}.

\section{Datasets}\label{sec:datasets}
Our stretched-grid models are flexible in terms of spatial resolution, and we utilize this property by training them on datasets of various resolutions. The temporal resolution is fixed at 6 hours.
For stage A, we use the ERA5 reanalysis \citep{hersbach2020} with training period 1979-01-01T00Z to 2021-12-31T18Z (43 years) and keep the year 2022 for validation. In total, 97 variables are input to the model and 87 variables are predicted, whereas 3 are diagnostic. We include 15 single level variables and 6 variables at 12 pressure levels. We refer to Table~\ref{tab:vars} for an overview of the variables. We use the variable weights from \citep{lang2024}, and $w_p=1$ for variables that were not included in that work.

\begin{figure*}
    \centering
    \includegraphics[width=0.9\textwidth]{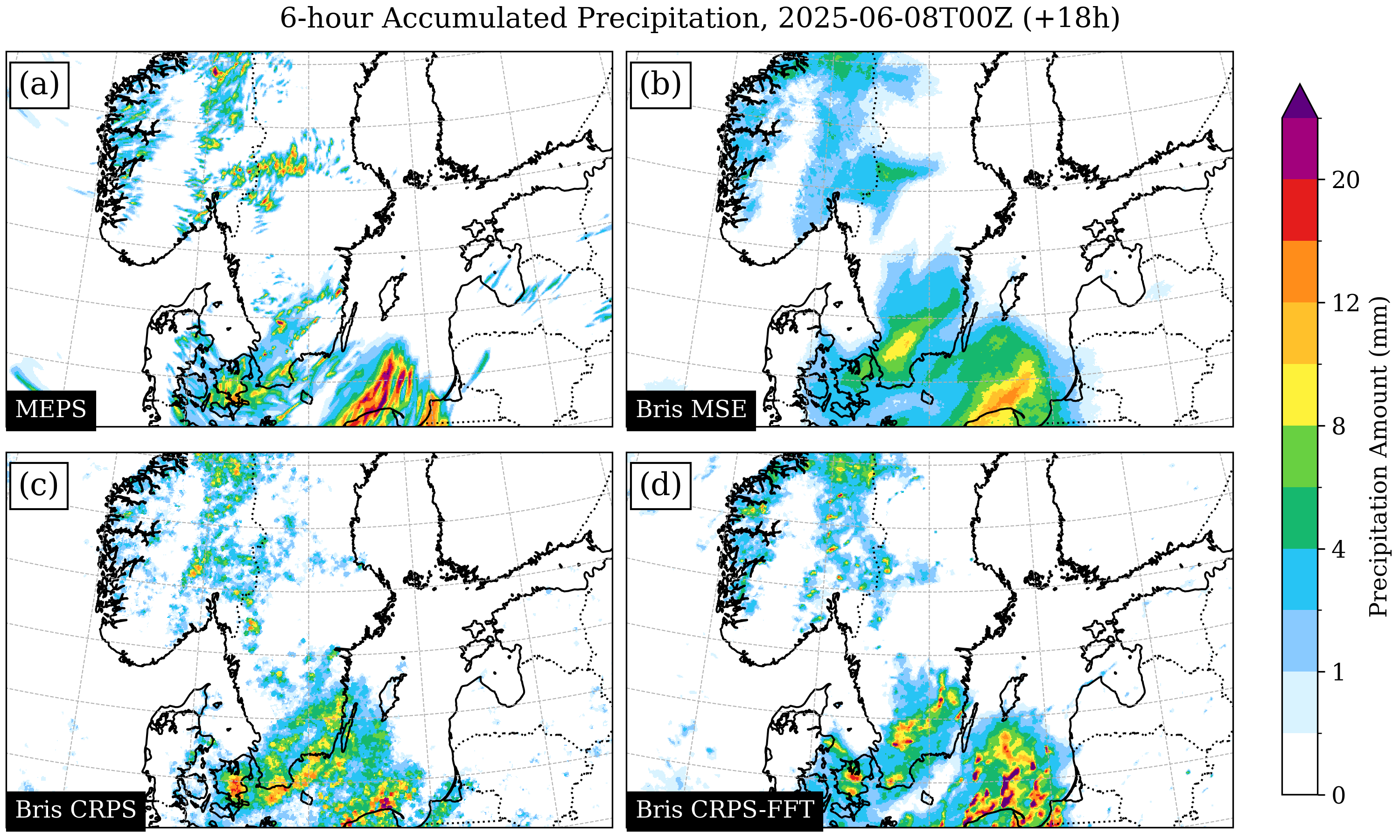}
    \caption{6-hour accumulated precipitation fields over parts of the Nordics at 2.5 km horizontal resolution. Each panel represents a different forecasting model: (a) The control member of the NWP modelling system \acrshort{meps}. (b) Bris MSE, as described in \citep{nipen2026}. (c) Bris CRPS, trained with point-wise \acrfull{crps} only. (d) \acrshort{bris} trained with point-wise and spectral \acrshort{crps}. Forecasts are initialized at 2025-06-08T00Z and have a lead time of +18 hours.}
    \label{fig:fields}
\end{figure*}

\begin{figure*}
    \centering
    \includegraphics[width=0.9\textwidth]{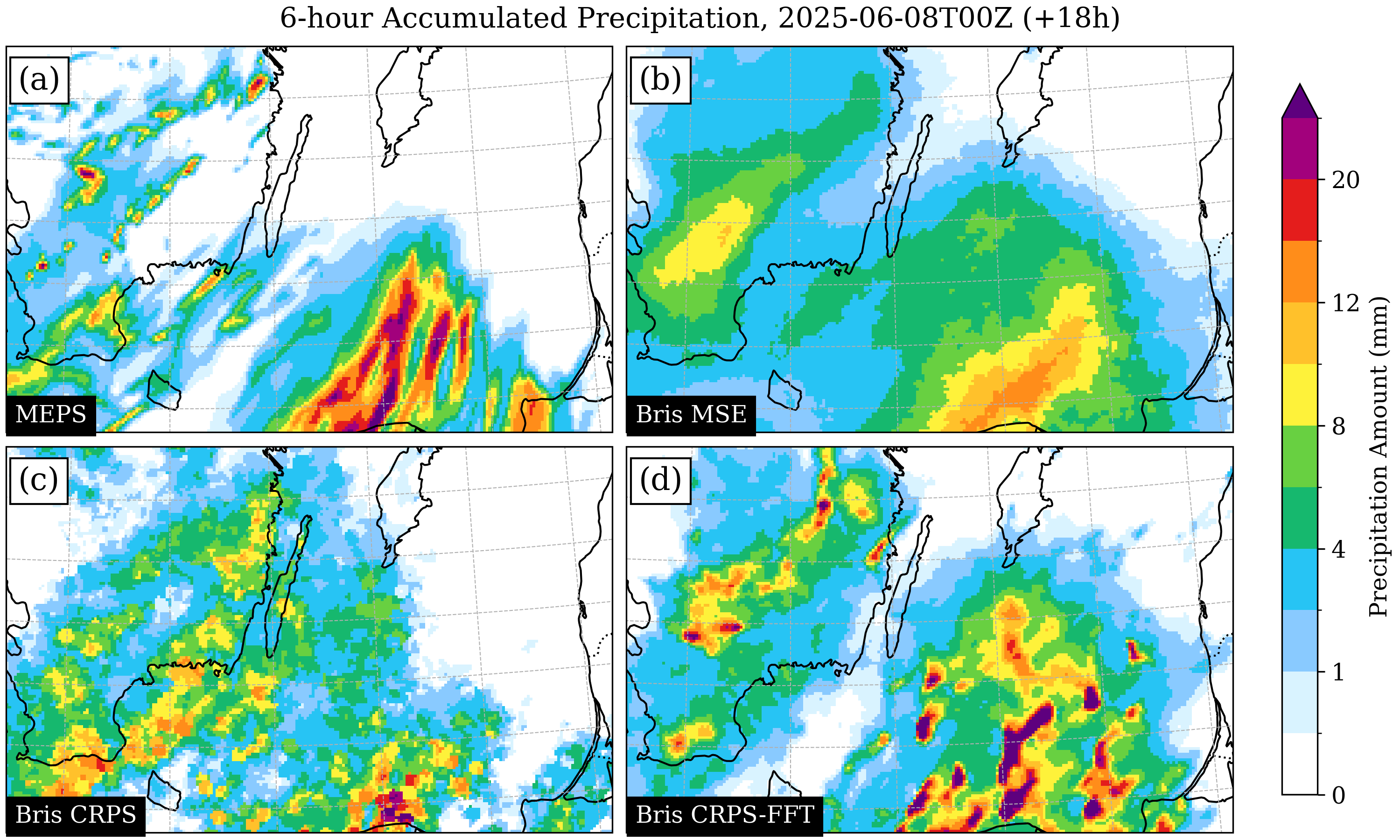}
    \caption{Same as Fig.~\ref{fig:fields}, but zoomed in over the southern Baltic Sea.}
    \label{fig:fields_zoom}
\end{figure*}

\begin{figure*}
    \centering
    \includegraphics[width=\textwidth]{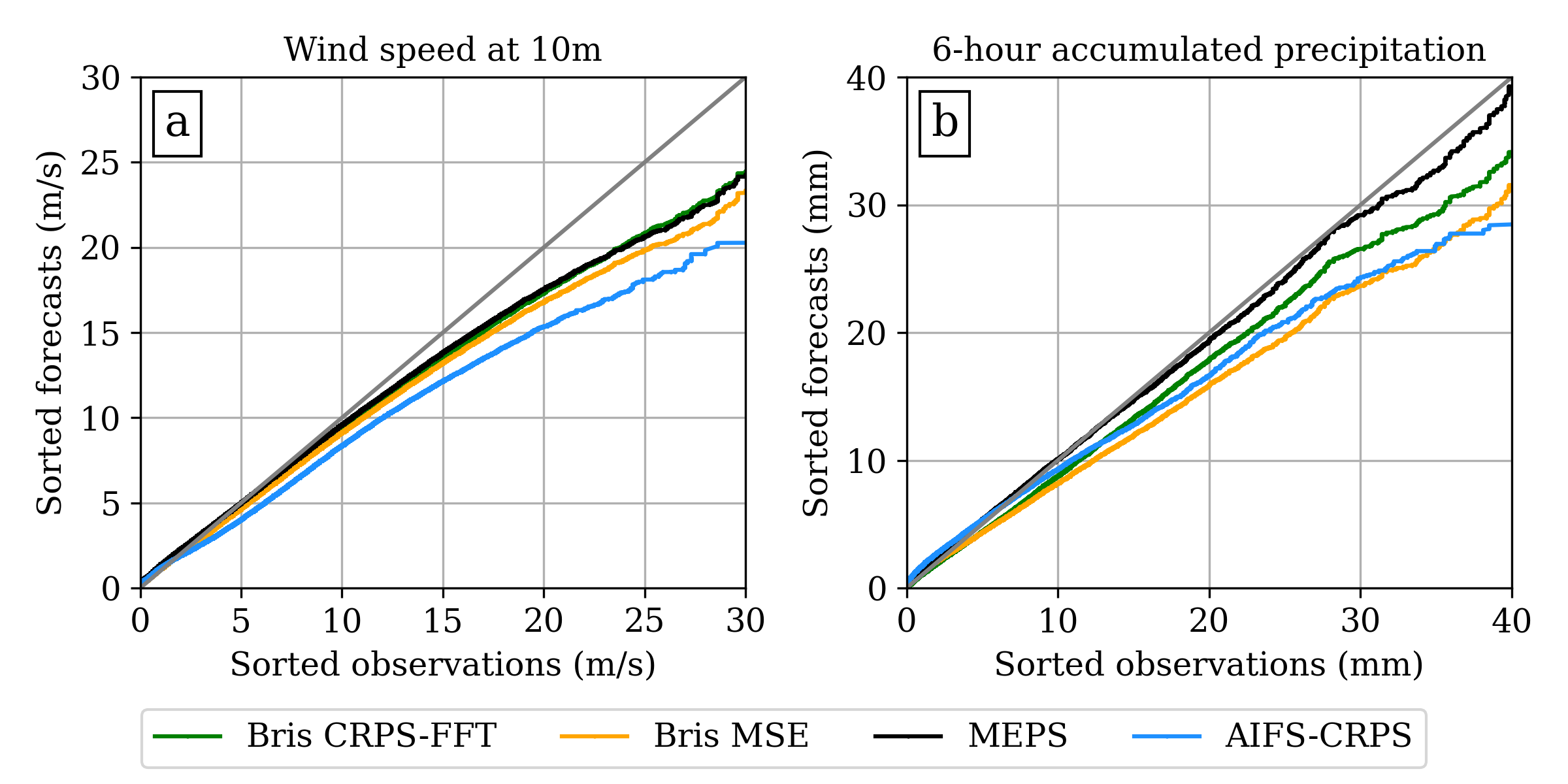}
    \caption{Quantile-quantile (QQ) plot aggregated over lead times 24 h--60 h. Forecasts are plotted as a function of observations, with the ideal case (gray line) being when the forecasts equal the observations. (a) is 10 m wind speed and (b) is 6-hour accumulated precipitation.}
    \label{fig:qq}
\end{figure*}

In stage B, the \acrfull{ifs}, regridded to the ERA5 grid, was used. The training period here is 2016-01-01T00Z to 2021-12-31T18Z (6 years), and we use the same validation period as for stage A.

Our stretched-grid formulation (stage C) requires both a global and a regional dataset, where the variables are consistent with the pre-training. Here we use the analysis of \acrshort{meps} over the Nordic domain and \acrshort{ifs} elsewhere. The 6-hour accumulated precipitation fields are retrieved from the first 6 hours of the control run forecasts. The training period here is from 2020-02-05T00Z to 2025-03-31T18Z ($\sim$ 5.2 years), and we keep 2025-04-01T00Z to 2026-04-08T18Z for testing. We trim the boundaries of \acrshort{meps} 125 km in all directions to avoid the model learning from data in the relaxation zone of the NWP analyses.

All datasets used in training and for inference initialization correspond to the control analysis, and are thus deterministic. We apply max normalization for surface geopotential, no normalization on the cloud coverage variables and forcings, and standard normalization on the remaining variables. A \acrshort{relu} bounding \citep{nair2010} was applied for precipitation to avoid negative values, while a hard tanh bounding \citep{glorot2011} was applied for cloud coverage variables to keep the range between 0 and 1.

\section{Results and discussion}\label{sec:results}
This section evaluates the quality of the weather forecasts from the perspective of a public weather forecast provider. To achieve this, we focus on surface variables of high relevance to end users, including air temperature at 2 m, wind speed at 10 m, 6-hour accumulated precipitation, and mean sea-level pressure. Mean sea-level pressure is selected as a representative synoptic-scale variable because direct observations are available and, over the Nordic region, it is strongly correlated with other large-scale atmospheric fields, such as 500-hPa geopotential height ($z_{500}$). We divide our verification into single-member and probabilistic evaluation.

%In Sec.~\ref{sec:single}, we perform a single member evaluation through\acrfull{qq} plots, in addition to the spatial properties of the fields through the \acrfull{dct} and inspection of the fields themselves. For probabilistic evaluation (Sec.~\ref{sec:prob}), we look at the \acrfull{crps}, spread-skill relation, and \acrfull{bss}.

With the exception of the visual analysis, all forecasts were verified against measurements from 254 synoptic weather stations throughout Norway. Forecasts were interpolated bilinearly to the observation points from the 4 nearest grid points in the model. As temperature is strongly related to altitude, we adjusted the interpolated temperatures by assuming a constant lapse rate of $6.5^{\circ}$C/km. This was applied to the altitude mismatch between the bilinearly interpolated altitude field and the station altitude, as has been done in \cite{nipen2026}.

\subsection{Single-member evaluation}\label{sec:single}
In this section, we evaluate the characteristics of the forecasted fields from a single ensemble member. We use the \acrshort{meps} control member and pick a random ensemble member from \acrshort{bris} and AIFS-CRPS. We also include Bris MSE in this comparison, to highlight the benefits of a probabilistic model.

\begin{figure*}
    \centering
    \includegraphics[width=\textwidth]{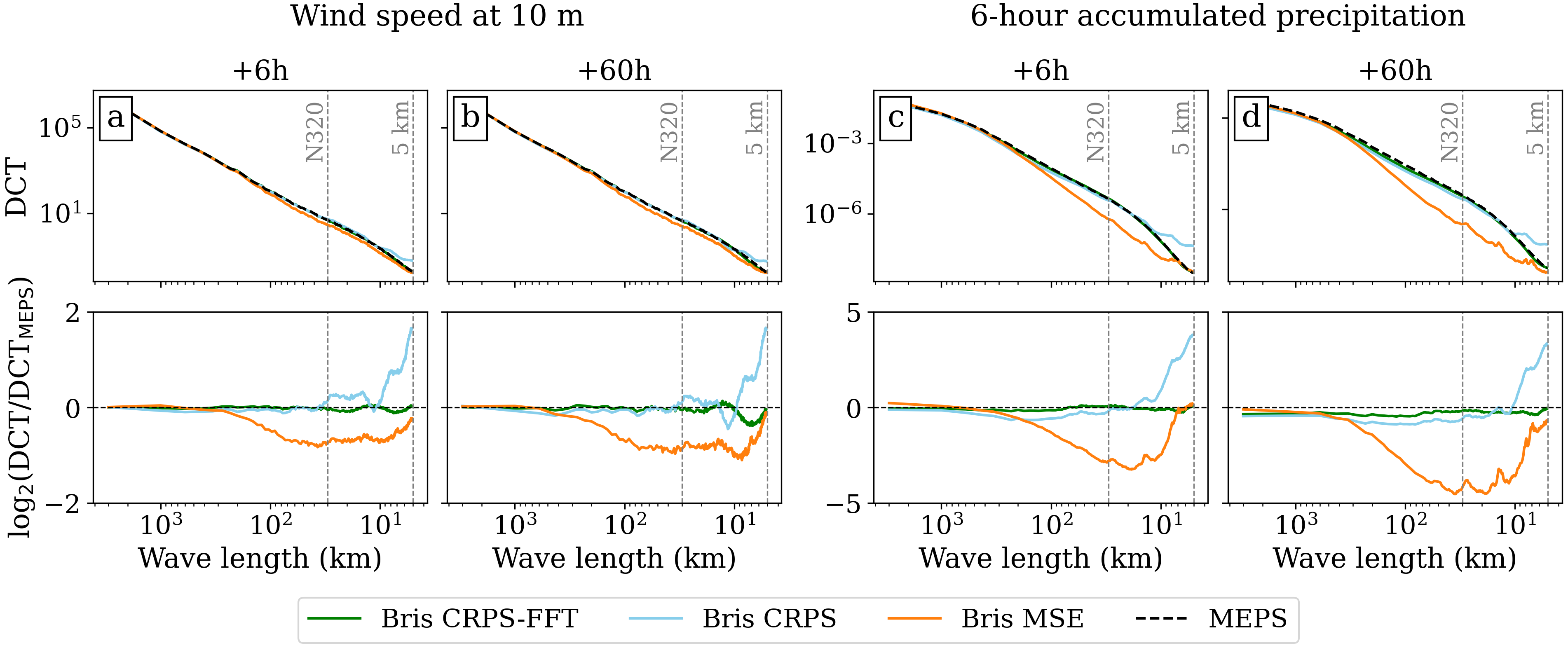}
    \caption{Upper row: Discrete cosine transform (DCT) power spectra as a function of wavelength for wind speed at 10 m for lead times +6 h (a) and +60 h (b), and for 6-hour accumulated precipitation for lead times +6 h (c) and +60 h (d). Lower row: the corresponding base-2 logarithmic power ratio with respect to \acrshort{meps}. The dashed lines mark the approximately 31 km N320 resolution and the 5 km Nyquist wavelength. The x-axis is flipped such that the wavelengths increase to the left. The power spectra are averaged from 2025-06-01T00Z to 2025-06-30T18Z.}
    \label{fig:dct}
\end{figure*}

Fig.~\ref{fig:fields} compares the 6-hour accumulated precipitation fields for an event initialized on 2025-06-08T00Z with a lead time of +18 h, for MEPS, Bris MSE, Bris CRPS, and \acrshort{bris}. MEPS tends to produce banded precipitation structures that reflect the dynamics of rain showers. It can also represent sharp features such as localized heavy precipitation. We treat the MEPS field as the reference here. In contrast, Bris MSE smooths out unpredictable events, such as heavy rainfall, which is evident from the overly smooth field structures. This smoothing results in a loss of detail, making it difficult to distinguish between sparse, localized showers and widespread light rain. Bris CRPS, on the other hand, demonstrates an improved ability to capture localized events but introduces substantial small-scale noise, which reduces the spatial coherence of such events.

\acrshort{bris} strikes a better balance by providing sharp features while preserving more spatial correlation. Although some residual small-scale noise is still present, it is less pronounced than in Bris CRPS. The spatial coherence may be further improved by tuning the spectral loss to more strongly constrain small-scale variability and by training on a larger dataset. Fig.~\ref{fig:fields_zoom} provides a zoomed-in view of the rain showers over the southern Baltic Sea, clearly highlighting the differences in precipitation forecasts among the models.

\begin{figure*}[htbp]
    \centering
    \includegraphics[width=0.8\textwidth]{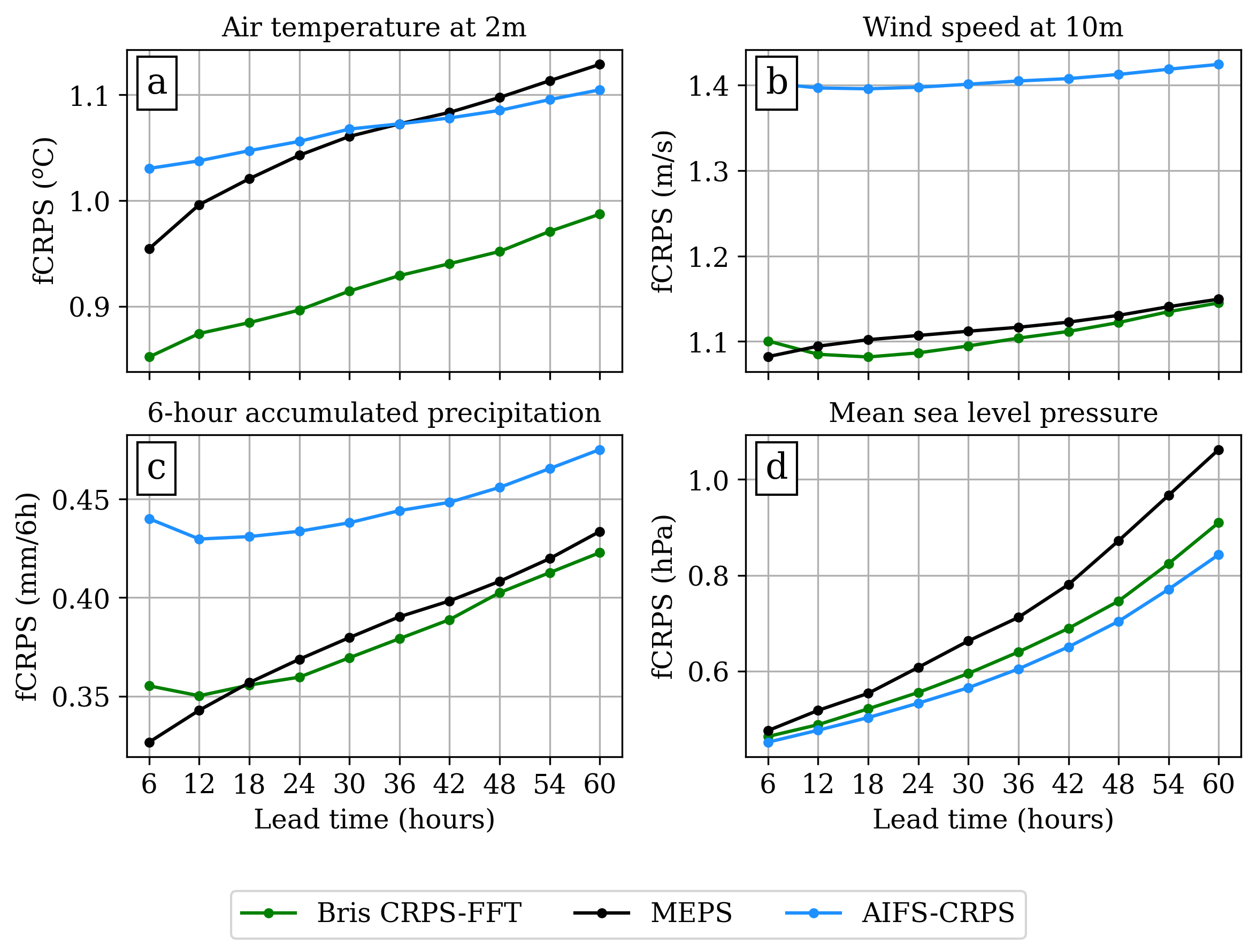}
    \caption{\acrfull{fcrps} plotted as a function of lead time for 2 m temperature (a), 10 m wind speed (b), 6-hour accumulated precipitation (c), and mean sea-level pressure (d). Lower score is better. Note that the y-axis is magnified to highlight differences between the forecasts.}
    \label{fig:crps}
\end{figure*}

Fig.~\ref{fig:qq} shows a \acrfull{qq} plot for 6-hour accumulated precipitation and 10 m wind speed, where sorted observations are plotted against sorted forecasts to assess the similarity between their distributions. While \acrshort{bris} underestimates the frequency of larger precipitation events, it performs better than Bris MSE, highlighting the added value of training Bris with a probabilistic approach. AIFS-CRPS also underestimates precipitation due to the coarser resolution, and so does \acrshort{meps}, but is closer to the ideal line than \acrshort{bris}.

All models in this comparison underestimate wind speed, a bias that can partially be attributed to the verification of gridded model outputs against point measurements. Bilinear interpolation between grid points smooths out local extremes, such as high wind speeds, resulting in systematically lower interpolated values compared to actual point observations. However, the relative differences between the models are interesting. A comparison between AIFS-CRPS and \acrshort{bris} highlights the potential advantage of higher spatial resolution, with \acrshort{bris} exhibiting a more realistic distribution. Furthermore, \acrshort{bris} shows notable improvement over Bris MSE in predicting higher wind speeds. The goal for \acrshort{bris} is to replicate the distribution of the target, represented by the MEPS analysis. The reason why \acrshort{bris} produces higher values than MEPS remains unclear and warrants further investigation.

To evaluate the ability of \acrshort{bris} to represent spatial variability across different scales, we analyze the power spectra of the forecasts. The spectra are computed using the \acrfull{dct} \citep{makhoul1980} instead of the standard \acrfull{fft} to mitigate boundary-related artifacts. The \acrshort{dct} is used only for evaluation; the \acrshort{fft} is used during training because it has a similar effect at lower computational and memory cost.

Figure~\ref{fig:dct} presents the \acrshort{dct} power spectra for wind speed and 6-hour accumulated precipitation at lead times of +6 h and +60 h for \acrshort{bris}, Bris CRPS, Bris MSE, and MEPS (upper row). To highlight the differences between the models, the logarithmic difference with respect to \acrshort{meps} is also shown (lower row), with MEPS treated as the reference.

The wind speed spectra show that \acrshort{bris} remains the closest to \acrshort{meps} at both +6 h and +60 h, with only a modest loss of power at the shortest wavelengths by +60 h. Bris CRPS is also close to \acrshort{meps} at larger scales, but develops a clear excess of small-scale power, strongest below about 10 km. In contrast, Bris MSE is systematically too smooth, with too little power from roughly mesoscales down to the shortest resolved wavelengths for both +6 h and +60 h.

For precipitation, \acrshort{bris} again agrees best with \acrshort{meps}, although it shows a slight underestimation of power that becomes more visible at +60 h. Bris CRPS is reasonably close at larger scales, but exhibits a pronounced build-up of excessive small-scale power at the shortest wavelengths, as for wind speed. Bris MSE has too little power across most scales, with the largest deficit at mesoscales and a substantially stronger underestimate at +60 h than at +6 h.

\subsection{Probabilistic evaluation}\label{sec:prob}
To evaluate the probabilistic properties of the forecasts, we generate 30 ensemble members for the entire verification period (2025-04-01T00Z to 2026-03-31T18Z) and compare with \acrshort{meps}, which runs operationally with 30 members. As \acrshort{meps} only provides forecasts up to 60 hours into the future, we focus the comparison on short-range forecasts. Additionally, we compare the model to 30 ensemble members generated by AIFS-CRPS \citep{lang2026} at a coarser resolution of 31 km, in order to assess the added value of using a higher-resolution \acrshort{ddm}. It is important to note that, in this setup, AIFS-CRPS is initialized with the control analysis from IFS, rather than the perturbed analyses from all IFS ensemble members, which is the operational configuration. We do this to run the model in a setup that is more comparable to \acrshort{bris}.

\begin{figure*}[htbp]
    \centering
    \includegraphics[width=0.8\textwidth]{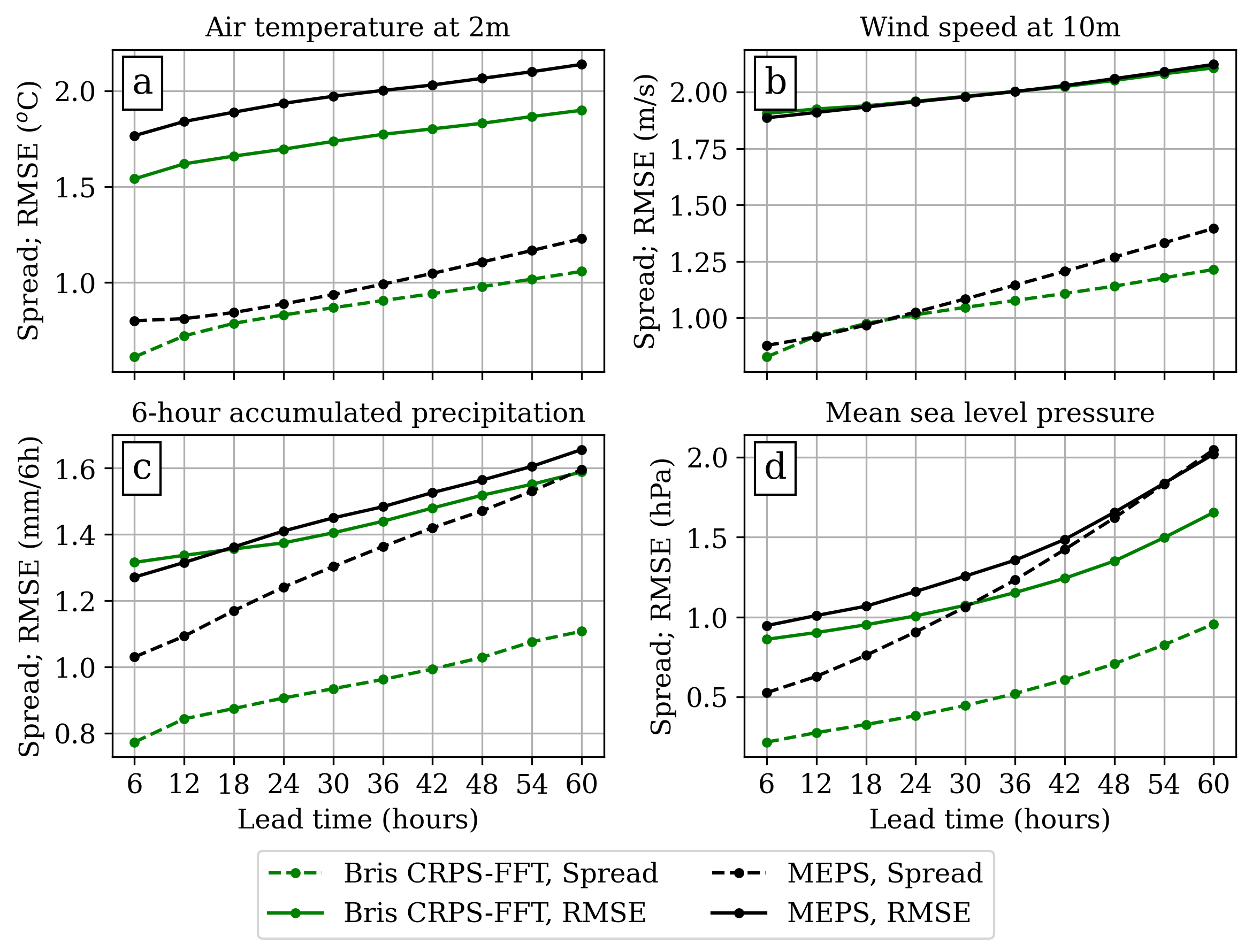}
    \caption{Spread (dashed lines) and \acrfull{rmse} of the ensemble mean (solid lines) as a function of lead times for 2 m temperature (a), 10 m wind speed (b), 6-hour accumulated precipitation (c), and mean sea-level pressure (d). Overlapping spread and \acrshort{rmse} curves give a perfect spread-skill ratio. Note that the y-axis is magnified to highlight differences between the forecasts.}
    \label{fig:ssr}
\end{figure*}

To evaluate the probabilistic forecast skill, the \acrshort{crps} metric is a natural starting point, as it is a proper scoring rule and the model is trained to optimize a closely related quantity. For this analysis, we use fair \acrshort{crps} \citep{ferro2014} (\acrshort{fcrps}), using Eq.~\ref{eq:afcrps} with $\varepsilon=0$. Figure~\ref{fig:crps} shows the \acrshort{fcrps} as a function of lead time, based on forecasts at the locations of 254 synoptic weather stations in Norway.

For temperature, we observe a clear improvement at all lead times, varying between 11\% and 15\% with an average improvement of 13\% compared to \acrshort{meps}. This is likely due in part to the \acrshort{meps} model's inability to retain analysis increments forward in time. That is, \acrshort{meps} temperature analyses are significantly more accurate than its forecasts, as discussed in \citep{nipen2026}.

For wind speed, \acrshort{bris} provides a slightly lower \acrshort{fcrps} compared to \acrshort{meps} for all lead times after 6 hours, with an average improvement of around 0.8\%. Both high-resolution models outperform AIFS-CRPS. The lower performance of AIFS-CRPS is expected, given the difference in spatial resolution, making it challenging to accurately represent 10 m wind speeds in mountainous regions.

For precipitation, \acrshort{bris} performs slightly worse than \acrshort{meps} at lead times up to 12 hours, but achieves slightly better performance at longer lead times, with an average improvement of around 0.6\%. The difference for short lead times is likely due to \acrshort{meps} being initialized from an ensemble of perturbed initial states, capturing initial state uncertainty, whereas \acrshort{bris} is initialized solely from the analysis of the \acrshort{meps} control run. This highlights a potential area for improvement for \acrshort{bris}.

For mean sea-level pressure, \acrshort{bris} outperforms \acrshort{meps} for all lead times, with the difference increasing rapidly from 3\% at lead time +6 h to 15\% at lead time +60 h. Notably, this is the only variable presented here where AIFS-CRPS performs better than the high-resolution models, which reflects the fact that it shows higher large-scale forecast skill than the higher-resolution models. In general, we observe that the models performing well on synoptic scales like \acrshort{mslp} are not the same models that perform best on variables like precipitation and wind speed, see also Figs.~\ref{fig:r1} and \ref{fig:area_weighting}.

We evaluate the ensemble spread-skill relationship by plotting \acrfull{rmse} of the ensemble mean and spread in the same panels (Fig.~\ref{fig:ssr}). We follow \cite{fortin2014} and define spread as the root mean ensemble sample variance, and skill as the RMSE of the ensemble mean, taking into account the sampling error of the ensemble mean of a finite sized ensemble. For an ideal spread-skill ratio, the \acrshort{rmse} (solid lines) and spread (dashed lines) should overlap completely. However, since we verify against observations, the \acrshort{rmse} of the model is expected to be higher than the spread. This is because the model grid points represent a larger spatial scale than the observations, and the observations are therefore not representative of the whole grid cell.

\acrshort{bris} has less spread than \acrshort{meps} for all four key surface variables presented here. A possible explanation of this is that \acrshort{bris} has been trained against the control analysis of \acrshort{meps}. The ensemble therefore aims to represent the uncertain evolution of the control analysis. It is important to note however, that because we verify against station observations, forecast errors includes observation and representativeness errors that are not represented by the forecast ensemble \citep{yamaguchi2016}.
This also suggests a potential benefit of initializing \acrshort{bris} on an ensemble of analyses that can capture initial condition uncertainty. For \acrshort{meps}, mean sea-level pressure forecast skill is comparatively lower, reflecting the limitations of short-range limited-area models in representing synoptic-scale features that are better captured by global models.

Despite having too little spread, the \acrshort{rmse} of the ensemble mean for \acrshort{bris} is in general better than for \acrshort{meps}, except for wind speed, where they are nearly identical. \acrshort{rmse} is highly correlated with \acrshort{fcrps} for all variables, as expected.

\begin{figure*}
    \centering
    \includegraphics[width=\textwidth]{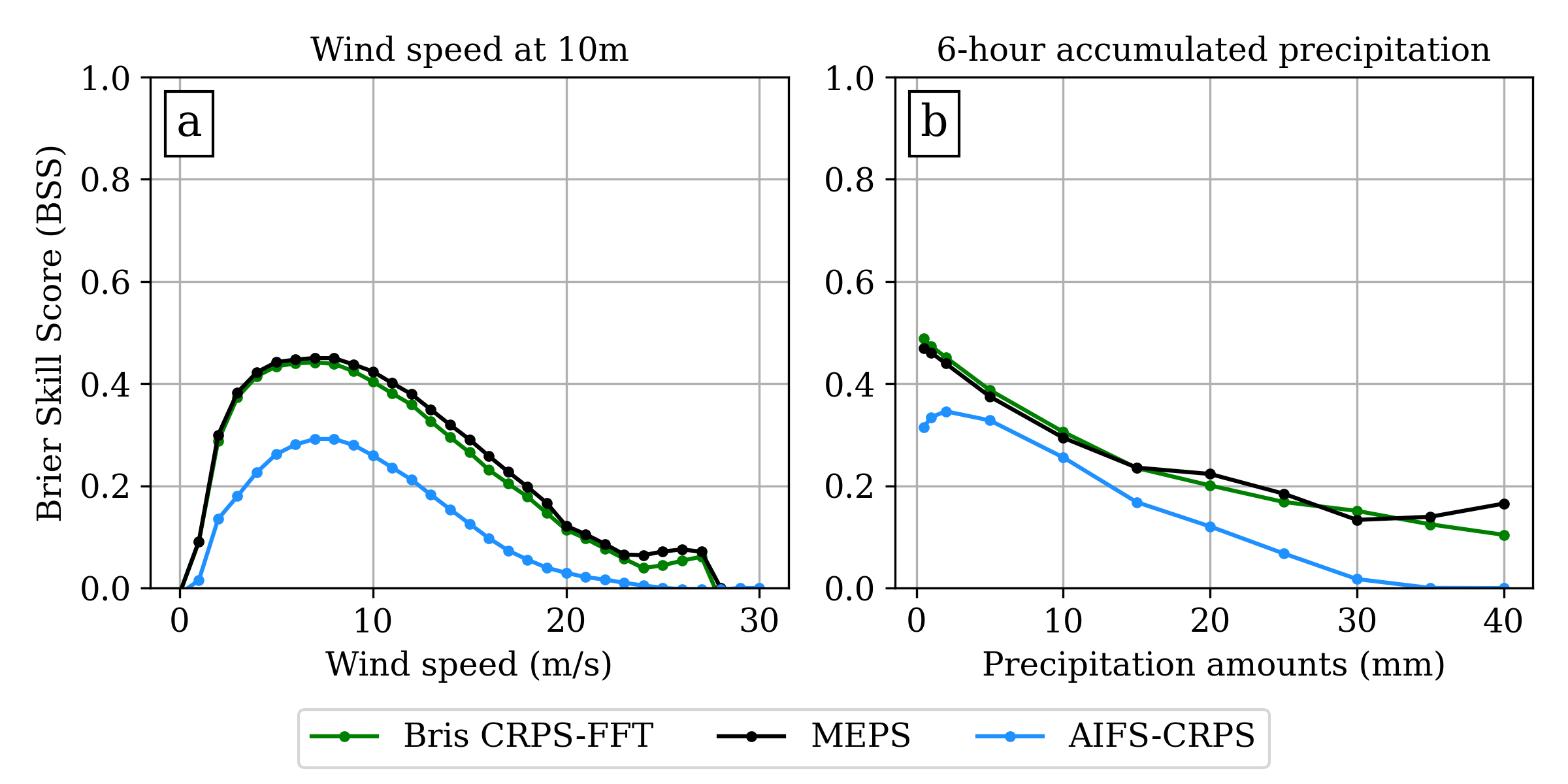}
    \caption{\acrfull{bss} as a function of thresholds for 10 m wind speed (a) and 6-hour accumulated precipitation (b). Higher score is better.}
    \label{fig:bss}
\end{figure*}

In Fig.~\ref{fig:bss}, we have plotted the \acrfull{bss} \citep{brier1950} as a function of thresholds for precipitation and wind speed. For wind speed, \acrshort{bris} performs similarly or slightly worse for all thresholds, despite performing better at \acrshort{fcrps}. Weak winds are greatly overrepresented in the distribution and are likely to dominate \acrshort{fcrps}. For precipitation, \acrshort{bris} and \acrshort{meps} perform similarly.

\subsection{Wind case: Storm Dave}\label{sec:dave}

Dave was an extreme weather event that affected southern Norway on 5 April 2026. The system developed rapidly after forming over the Atlantic near Ireland, before moving into the North Sea. Norwegian authorities issued red warnings for parts of the coast. Since \acrshort{bris} predicts 10 m wind speed (10ff), we focus on mean wind speed rather than wind gusts. The strongest observed mean wind speeds during Dave occurred at exposed mountain and coastal stations, reaching 37.9 m/s at Folgefonna Skisenter and 33.9 m/s at Eigerøya.

To assess how well \acrshort{bris} captured the strong winds, we use an approach similar to that applied in operational warning assessment. Since the magnitude of the wind is more important here than the exact timing of the peak, we evaluate the 24 h maximum wind speed, thereby reducing the impact of timing errors. Because the forecasts have 6-hourly temporal resolution, this maximum is computed over four forecast valid times. For Dave, these valid times span 2026-04-05T06Z to 2026-04-06T00Z.

\begin{figure*}[htbp]
    \centering
    \includegraphics[width=\textwidth]{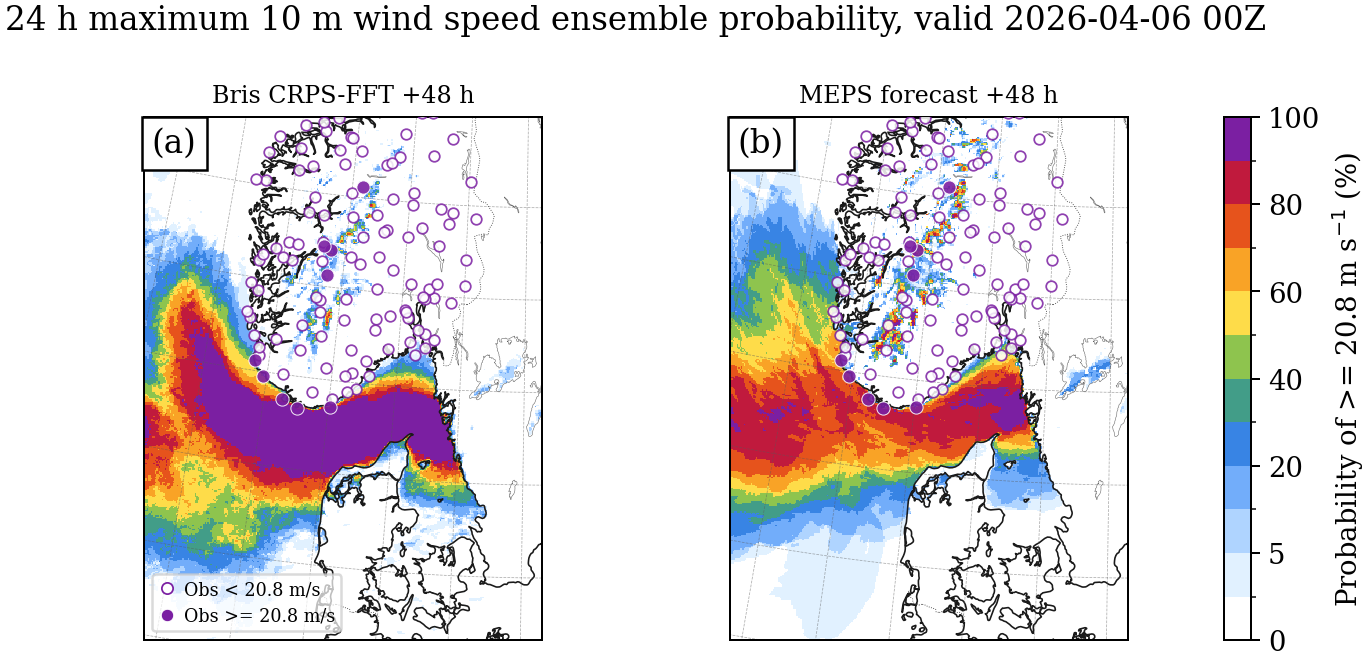}
    \caption{24 h ensemble probability of 10 m wind speed exceeding 24 m/s during the high-impact Storm Dave over southern Norway. Panel (a) shows the Bris CRPS-FFT forecast and panel (b) the corresponding MEPS forecast, both issued 48 h before the event. Probabilities are calculated from the maximum wind speed over the period from 2026-04-05T06Z to 2026-04-06T00Z. Observation points indicate whether the observed 24 h maximum wind speed was below 24 m/s (open circles) or at least 24 m/s (filled circles).
}
    \label{fig:dave}
\end{figure*}

Figure~\ref{fig:dave} shows the probability that the 24 h maximum 10 m wind speed exceeds the strong-gale threshold of 20.8 m/s over southern Norway during Storm Dave. Panel (a) shows the \acrshort{bris} forecast and panel (b) the corresponding \acrshort{meps} forecast, both issued 48 h before the event. The overlaid observation points provide the best available reference, indicating whether the observed 24 h maximum wind speed was below or above the strong-gale threshold.

Both \acrshort{bris} and \acrshort{meps} predict high probabilities of strong-gale winds along the exposed southern Norwegian coast and in high-terrain areas. The \acrshort{bris} forecast in panel (a) displays a larger, more spatially coherent area of high probability, whereas the \acrshort{meps} forecast in panel (b) has a more fragmented probability pattern. The broader high-probability region in \acrshort{bris} may indicate overconfidence, consistent with indications from the spread--skill analysis. The binary observations show that strong-gale winds occurred at several stations, including stations close to areas with elevated forecast probability. However, both systems assign relatively low probabilities at some coastal and mountainous stations where strong-gale winds were observed, illustrating the difficulty of forecasting local wind extremes at a 48 h lead time.

The strongest observed winds occurred between two 6-hourly forecast valid times, which likely contributed to the forecasts not reaching the most extreme observed values. Nevertheless, the forecast winds are still strong and indicate that \acrshort{bris} is able to represent substantial wind anomalies. A more complete assessment of its ability to capture extreme and high-impact wind events would require a longer test period with more extreme events and, preferably, forecasts with hourly temporal resolution. This is outside the scope of the present paper.

\subsection{Forecasting efficiency and scalability}

\acrshort{bris} performs a single-member 3-day forecast in around 30 seconds on an NVIDIA H200 \acrfull{gpu}, which is the hardware used for our in-house operational runs at MET Norway. The NVIDIA H200 \acrshort{gpu} has a maximum configurable \acrshort{tdp} of up to 600 W. Under this compute-device \acrshort{tdp} proxy, a single MEPS member forecast requires approximately 2.3 kWh, while a single \acrshort{bris} forecast requires approximately 5 Wh.

%%%%%%%%%%%%%%%
% CONCLUSIONS %
%%%%%%%%%%%%%%%
\section{Conclusion}\label{sec:conclusion}
In this article, we have presented \acrshort{bris}, a high-resolution probabilistic \acrshort{ddm} trained on point-wise and spectral \acrshort{crps}. The model provides forecasts with skill comparable to, and for some variables and metrics better than, the operational \acrshort{nwp} ensemble prediction system \acrshort{meps}. The model provides lower \acrshort{fcrps} for all lead times after +12 h across the key variables evaluated, with an average improvement of 10\% for mean sea-level pressure and 13\% for 2 m temperature.

We have shown that the spatial characteristics of single ensemble members are closer to those of \acrshort{nwp} models than to those of fields from an \acrshort{mse}-based \acrshort{ddm}. Incorporating terms into the loss function that assess the model's ability to represent different spatial scales was essential to produce coherent fields.

\begin{figure*}[p]
    \centering

    \includegraphics[width=0.8\textwidth]{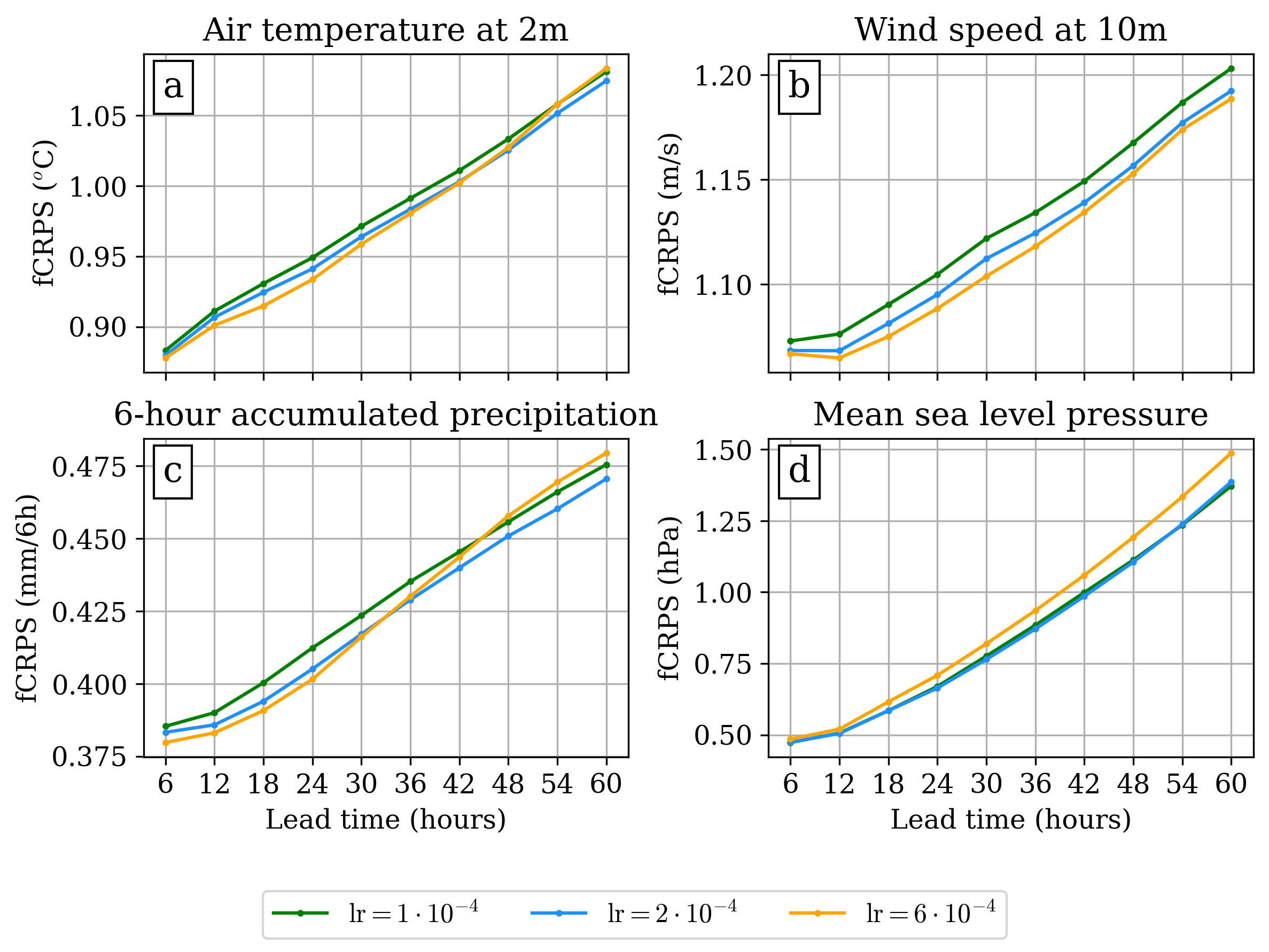}
    \caption{Results for tuning the learning rate in stage C1.
    \acrfull{fcrps} plotted as a function of lead time for
    2 m temperature (a), 10 m wind speed (b),
    6-hour accumulated precipitation (c), and mean sea-level
    pressure (d). Lower score is better.}
    \label{fig:r1}

    \vspace{1em}

    \includegraphics[width=0.8\textwidth]{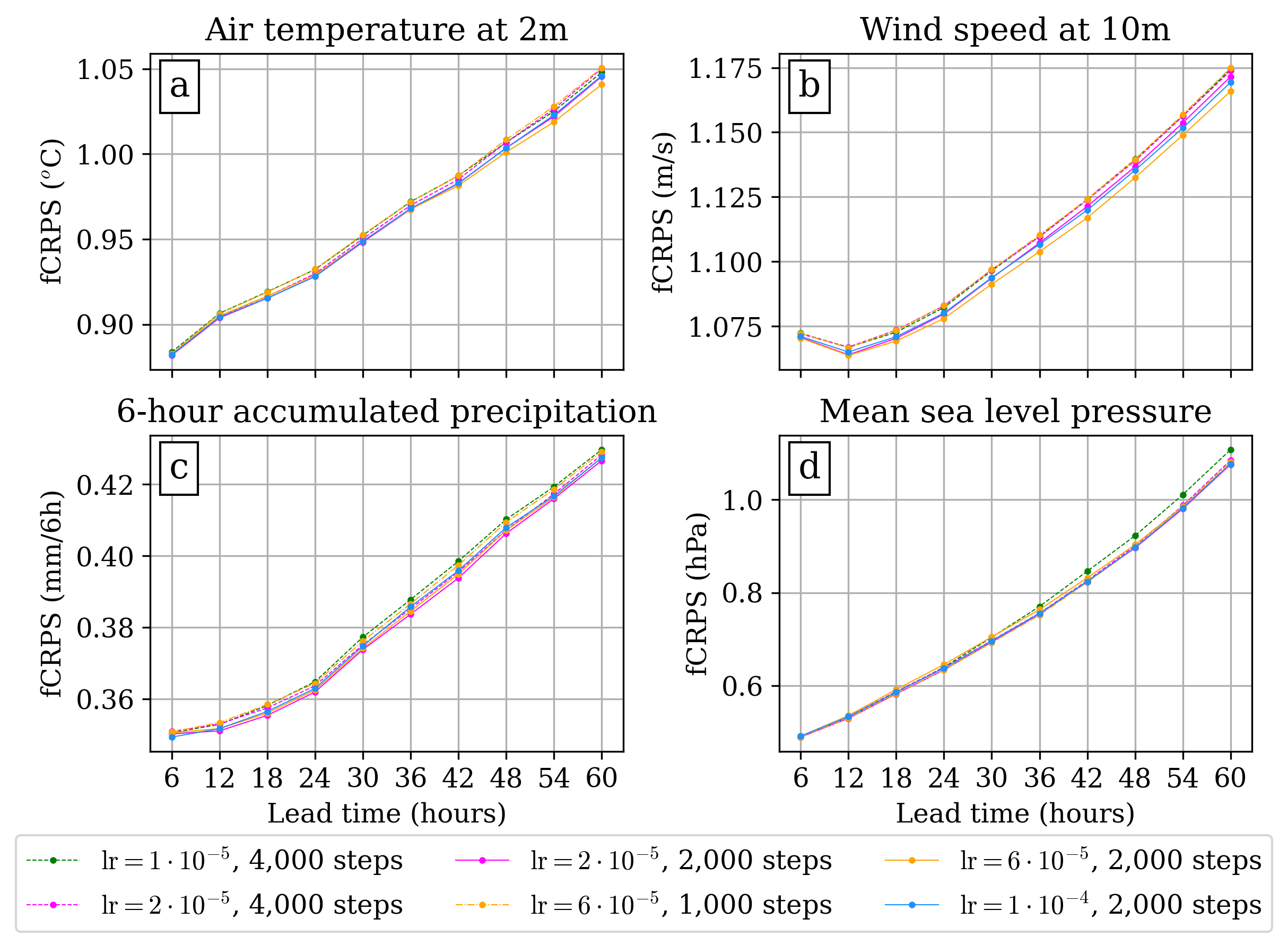}
    \caption{Results for tuning the learning rate and number of
    iterations for 12 h rollout in stage C2.
    \acrfull{fcrps} plotted as a function of lead time for
    2 m temperature (a), 10 m wind speed (b),
    6-hour accumulated precipitation (c), and mean sea-level
    pressure (d). Lower score is better. Learning rates are
    separated by line color and number of steps by line style.}
    \label{fig:r2}
\end{figure*}

The current model operates at 6-hour temporal resolution, which is too coarse for many applications. A recent work has explored increasing the temporal resolution to hourly forecasts using a temporal downscaler, better matching our high spatial resolution \cite{ingstad2026}.

We have observed that the model is overconfident with the current setup, providing too little ensemble spread and not spanning the range of possible outcomes. This can largely be attributed to the lack of initial-state perturbations, but even against the training dataset the spread is too low. For future work, we will include perturbed initial states when running the model, accounting for uncertainties in the initial state. Part of the lack of spread can be attributed to the model overfitting on the training data, in the sense that skill is better than for validation while spread stays the same. In future work, we will redo the model training with 40 years of the Copernicus Arctic Regional Reanalysis version 2 (CARRA2) dataset.

Issuing timely and reliable extreme-weather warnings is a core responsibility of \acrfull{nmhses}. Yet forecasting high-impact events remains challenging because such phenomena are short-lived, spatially localized, and sensitive to initial-condition uncertainty. High-resolution probabilistic guidance is therefore essential for meeting operational warning requirements. In this context, \acrshort{ddms} offer a promising pathway for next-generation extreme-weather prediction owing to their fast forecasting speed and capacity for efficient ensemble generation, enabling much larger ensembles and thereby improved sampling of the forecast probability distribution.

Furthermore, the stretched-grid design of \acrshort{bris} enables efficient operational deployment; the system has been running operationally at MET Norway four times a day since October 2025 and is used for internal purposes.

\subsection*{Code availability}

The training and inference checkpoints, together with the configuration files required to fine-tune and run the model, are publicly available in the Bris Forecaster repository on Hugging Face: \url{https://huggingface.co/met-no/bris-forecaster}.

\subsection*{Acknowledgments}
This work was funded by the Horizon Europe project WeatherGenerator (grant agreement 101187947). Computing and storage resources were provided by EuroHPC through the regular access calls EHPC-REG-2024R02-079 and EHPC-REG-2025R02-263. We thank Svante Henriksson, Bastien Francois and Boris Bonev for fruitful discussions about how to set up the loss objective.

\appendix

\section{Hyperparameter tuning} \label{sec:hyperparameter}
Extensive hyperparameter tuning was performed for the transfer learning and rollout of the stretched-grid model in stage C. The transfer learning is sensitive to learning rate: a value that is too small will make it hard for the model to adapt to the new domain, while a value that is too large can easily make the model forget what it learned in pre-training. A small learning rate can be compensated for to some degree by increasing the number of training iterations. This is also closely related to the area weighting, defining how much weight is put on the regional domain, which is kept at 0.23 as in \cite{nipen2026} unless otherwise stated.

The spectral weighting defines the relative weighting of the point-wise and spectral loss terms and is kept fixed at 0.1 unless otherwise stated.
We cannot afford to perform a complete grid search of the hyperparameters, but tune one at a time to gain insight into how each affects the model.

All tuning of the stretched-grid model was performed using a model with a training period from 2020-02-05T00Z to 2022-05-31T18Z (2.3 years), 2022-06-01T00Z to 2023-05-31T00Z for validation, and 10 ensemble members. Note that the test period used to assess the final model in Section~\ref{sec:results} is independent of the validation period used for hyperparameter tuning.

\subsection{Learning rate and number of iterations}
We experimented with different learning rates for both the transition from stage B to C1 and that from stage C1 to C2.

%\subsubsection{6-hour rollout}
This is where the regional domain is introduced, and the focus is to make the model adapt to the new domain without forgetting the synoptic dynamics it learned during pre-training. We transfer from the same pre-trained model and run for 10,000 steps in all the experiments. In Fig.~\ref{fig:r1}, we compare the effective learning rates $1\cdot10^{-4}$, $2\cdot10^{-4}$ and $6\cdot10^{-4}$. We observe that the lowest learning rate performs worst on all the fields evaluated here, except mean sea-level pressure, which is the field that changes the least during transfer learning. This indicates that the model has not yet learned all the high-resolution dynamics. The highest learning rate tends to be good at short lead times, but worse at longer lead times, most pronounced for 6-hour accumulated precipitation, indicating too little spread at longer lead times. A learning rate of $2\cdot10^{-4}$ performs best overall and was therefore used for the production runs.

%\subsubsection{12-hour rollout}
Getting the 12-hour rollout right in stage C2 is essential, as this is the stage at which the model learns to use its own predictions as input for subsequent steps. In Fig.~\ref{fig:r2}, we evaluate six learning-rate/step combinations: ($1\cdot10^{-5}$, 4,000), ($2\cdot10^{-5}$, 2,000), ($2\cdot10^{-5}$, 4,000), ($6\cdot10^{-5}$, 1,000), ($6\cdot10^{-5}$, 2,000) and ($1\cdot10^{-4}$, 2,000). The lowest learning rate performs worst in our verification, whereas the remaining configurations perform similarly. Among these, the configuration with learning rate $6\cdot10^{-5}$ and 2,000 steps gives the best wind speed results. We therefore use this configuration in the following experiments, although several alternatives perform comparably well.

\subsection{Area weighting}
Here, the goal is to assess which area weighting $\lambda_{\mathrm{r}}$ gives the best model. All the runs are based on the same pre-trained model, and we train for 10,000 iterations with a 6-hour rollout, learning rate $1\cdot 10^{-4}$ and 1,000 warm-up steps.

\begin{figure*}[p]
    \centering

    \includegraphics[width=0.8\textwidth]{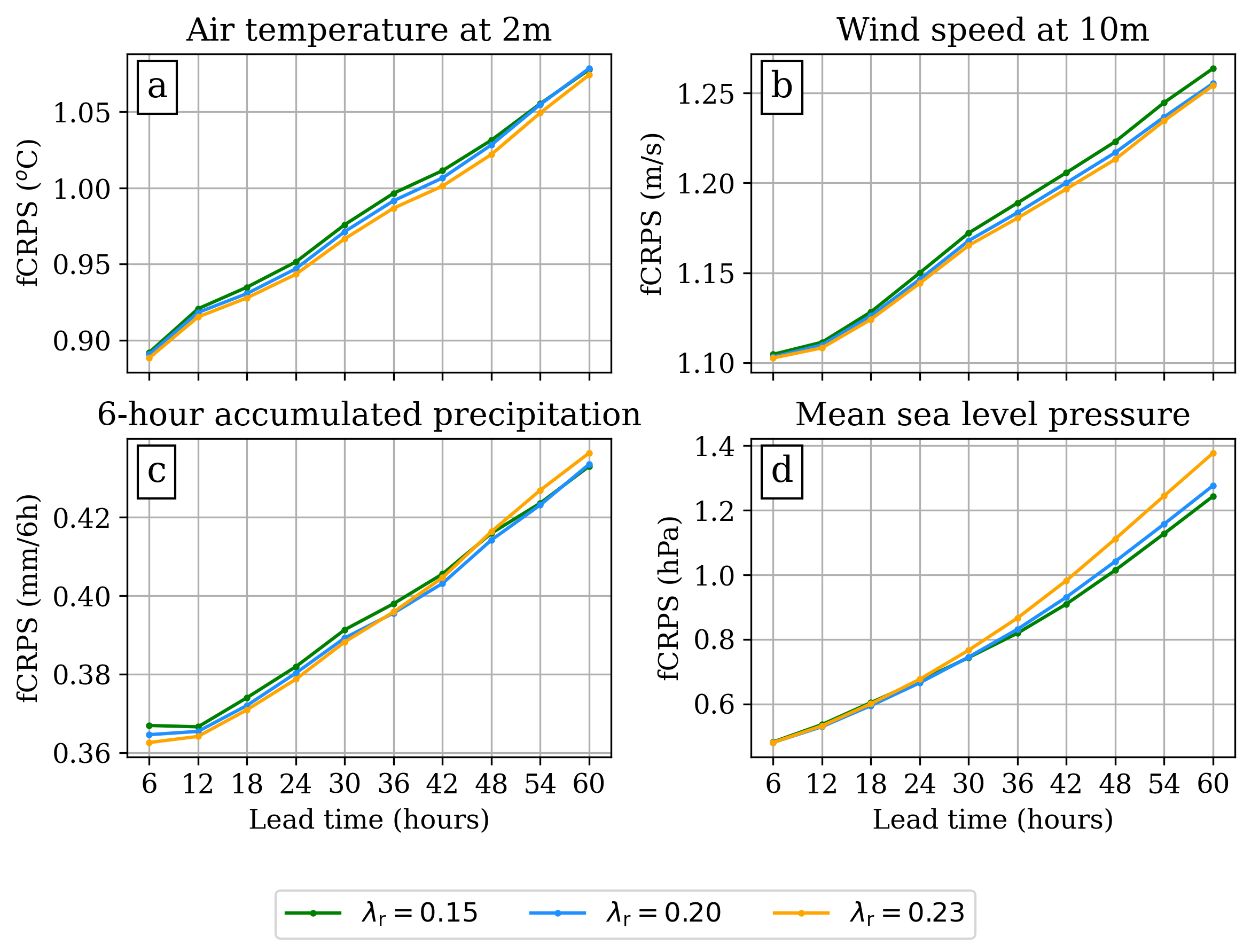}
    \caption{Results for tuning the area weighting in stage C1. \acrfull{fcrps} plotted as a function of lead time for 2 m temperature (a), 10 m wind speed (b), 6-hour accumulated precipitation (c), and mean sea-level pressure (d). Lower score is better.}
    \label{fig:area_weighting}

    \vspace{1em}

    \centering
    \includegraphics[width=0.8\textwidth]{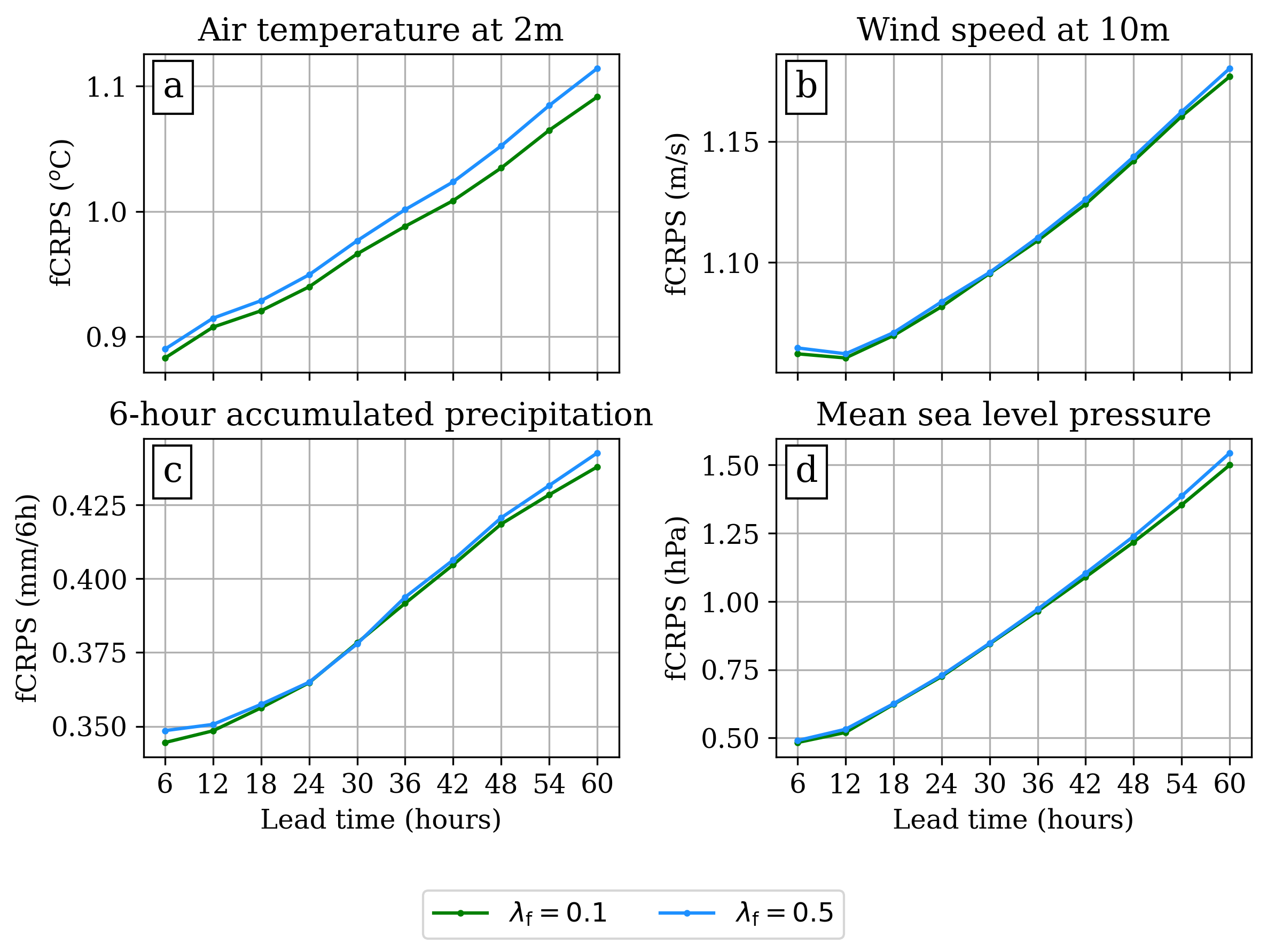}
    \caption{Results for tuning the spectral weighting. \acrfull{fcrps} plotted as a function of lead time for 2 m temperature (a), 10 m wind speed (b), 6-hour accumulated precipitation (c), and mean sea-level pressure (d). Lower score is better.}
    \label{fig:spectral_weighting}
\end{figure*}

\begin{figure*}[htbp]
    \centering
    \includegraphics[width=0.8\textwidth]{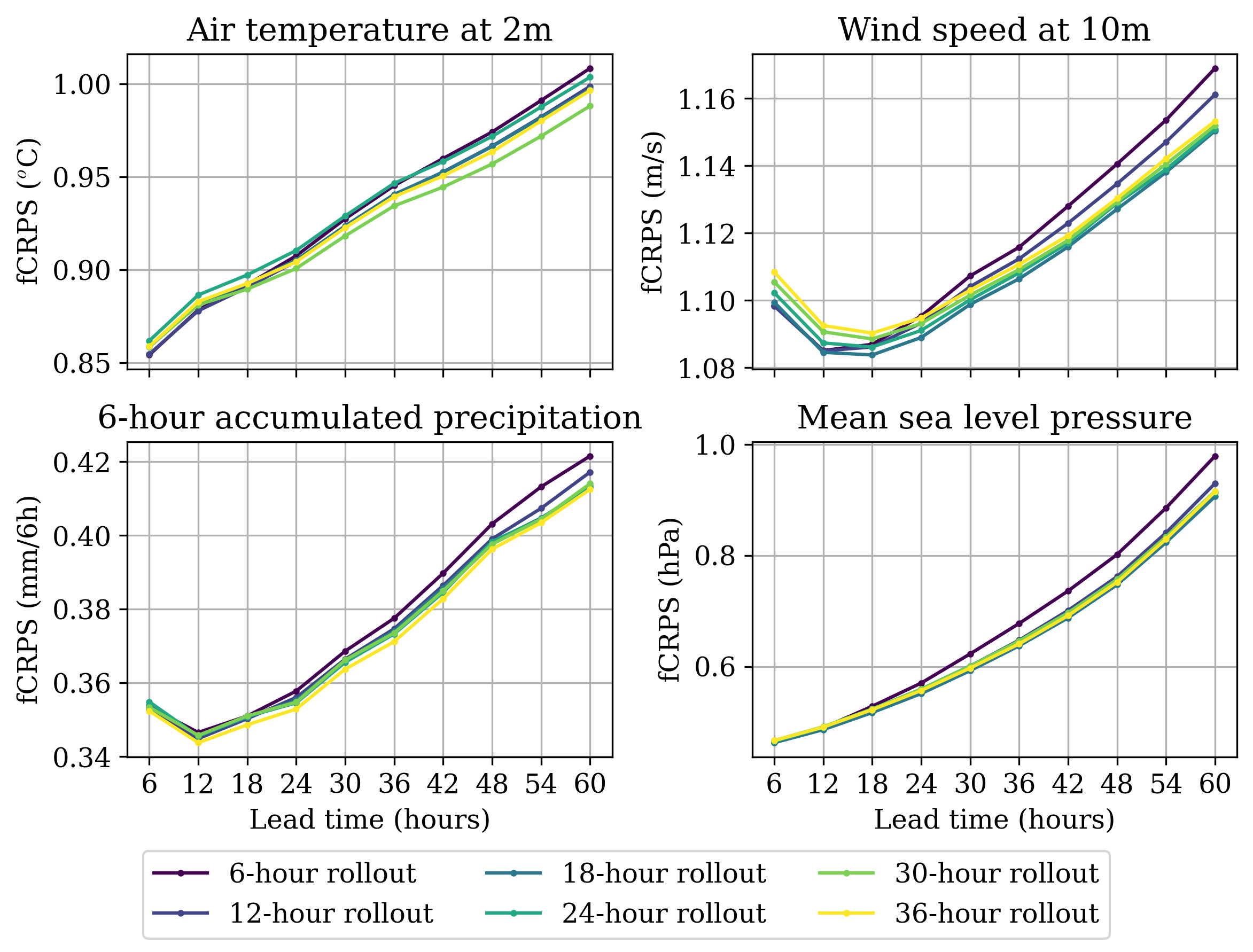}
    \caption{\acrfull{fcrps} plotted as a function of lead time for 2 m temperature (a), 10 m wind speed (b), 6-hour accumulated precipitation (c), and mean sea-level pressure (d). Lower score is better.}
    \label{fig:rollouts}
\end{figure*}

In Fig.~\ref{fig:area_weighting}, we compare $\lambda_{\mathrm{r}}=0.15, 0.20$ and $0.23$. We observe that the model with the lowest area weighting performs best for synoptic-scale variables such as mean sea-level pressure, but is worse for 2 m temperature, 10 m wind speed and 6-hour accumulated precipitation. This is likely because it has not learned the dynamics of \acrshort{meps} as well as the other models. The highest area weighting tested here, $\lambda_{\mathrm{r}}=0.23$, performs best overall, as in \cite{nipen2026}, and was selected for our production run.

\subsection{Spectral weighting}
The spectral weighting controls the trade-off between enforcing spatial coherence and minimizing the point-wise CRPS. If the spectral term is weighted too much, the model may learn undesirable shortcuts. For example, it may preserve spatial structure by simply replicating the initial state across forecast lead times, rather than producing dynamically evolving forecasts.

Figure~\ref{fig:spectral_weighting} compares the \acrshort{fcrps} for models trained with $\lambda_{\mathrm{f}}=0.1$ and $\lambda_{\mathrm{f}}=0.5$ for stage C1. The model with $\lambda_{\mathrm{f}}=0.1$ achieves a lower \acrshort{fcrps} than the model with $\lambda_{\mathrm{f}}=0.5$, particularly for 2 m temperature. Since $\lambda_{\mathrm{f}}=0.1$ also produces spatially coherent fields, it is used as a suitable compromise between point-wise accuracy and spatial consistency.

\section{Model evolution with rollout steps} \label{sec:rollout}
In this section, we examine how the \acrfull{fcrps} changes as the rollout length used during training is gradually increased, corresponding to forecast horizons from 6 to 36 hours. The results are evaluated over the test period defined above, using 10 ensemble members. The results are shown in Fig.~\ref{fig:rollouts}.

For some variables, such as 6-hour accumulated precipitation and mean sea-level pressure, performance improves monotonically with each additional rollout step. However, the improvement at short lead times is small after rollout step 3. For temperature at 2 m and wind speed at 10 m, the overall \acrshort{fcrps} does not consistently improve with increasing rollout length. The best-performing model for temperature at 2 m uses 5 rollout steps, while the best-performing model for wind speed at 10 m uses 3 rollout steps. Nevertheless, the model trained with 6 rollout steps shows the most favourable lead-time evolution and would likely perform better at longer forecast horizons. For short-range forecasting, we therefore see no clear benefit in increasing the rollout length beyond 6 steps, particularly since longer rollouts also incur higher computational and memory costs.

At first, it may seem counterintuitive that the \acrshort{fcrps} can increase when using more rollout steps, since this is essentially the quantity being optimised during training. However, an increase in \acrshort{fcrps} does not necessarily imply that the deterministic skill deteriorates. In our verification, the \acrshort{rmse} consistently improves as the number of rollout steps increases. In some cases, however, the ensemble spread decreases faster than the error, causing the forecasts to become increasingly overconfident. This leads to a higher \acrshort{fcrps}, despite improved \acrshort{rmse}. It is also important to note that a model can appear overconfident in verification without being overconfident during training, since the training forecasts have lower skill while exhibiting a similar spread.

%%%%%%%%%%%%%%%%
% BIBLIOGRAPHY %
%%%%%%%%%%%%%%%%
%\bibliographystyle{unsrt}
\bibliographystyle{ieeetr}
\bibliography{references.bib}  %%% Remove comment to use the external .bib file (using bibtex).


% \end{thebibliography}

\end{document}